# Nonlocal Optical Real Image Formation Theory

**Greyson Gilson**

Mulith Inc.
30 Chestnut Street # 32
Nashua, New Hampshire 03060-3361, USA
Email: greyson.gilson@mulithinc.com

## ABSTRACT

A nonlocal theory of optical real image formation is developed from the basic quantum physics linked to an optical real image formation apparatus. Optical real images are formed by photons. Photons are nonlocal quantum objects that exhibit wave-like properties and particle-like properties. Optical real image formation is nonlocal because at least two separated object points illuminated by a single photon are required for real image formation. When the distance between the illuminated points exceeds twice the wavelength of the light used two equi-amplitude plane waves propagate independently away from the object plane toward the imaging system. Propagation of photons away from a single point in the aperture plane does not occur. Photons with spatial frequencies that are within the passband of the imaging system pass through the imaging system. Each plane wave pair that propagates through the imaging system ultimately arrives at the image plane. At the image plane each plane wave pair overlaps to form an image of its associated pair of initial points on the object plane and thus contributes to real image formation. Object and image field depths are treated and found to be very different than they are ordinarily thought to be. Resolution limits related to real image formation are investigated. Reference distribution real image formation (RIF) is introduced. Real image formation with no fundamental resolution limit is predicted to be achievable by means of RIF.

## PLAN OF THE PAPER

## I. INTRODUCTION

A new theory of optical real image formation is developed from fundamental quantum physics[1] in this paper. In accord with the new theory, optical real images with a very large field depth (object and image) and no fundamental resolution limit can be formed. No diffraction limit exists.

Quantum objects, known as photons, are associated with optical real image formation. Photons cannot be observed directly; what is known of them comes from observing the results of their being either created or destroyed[2]. Nevertheless, photons display coexisting wave-like properties and particle-like properties. This coexistence is known as wave-particle duality[3].

As a principal ingredient, the unified theory of wave-particle duality includes the hypothesis that the optical differential equation is applicable to all quantum objects. Thus, the quantum amplitude $\Psi$ linked to a quantum object satisfies the differential equation

$$\nabla^2\Psi = \frac{\partial^2\Psi}{\partial t^2} \tag{1.1}$$

In this equation, $t$ is time, $\nabla^2$ is the Laplacian operator and the speed of wave propagation (unity – the speed of light) is the same for all quantum objects; the units used for time are the same as those used for distance.

Equation (1.1) is not uniquely associated with quantum objects. Thus, in addition to describing quantum amplitudes, $\Psi$ describes the complex amplitudes of classical electromagnetic theory.

Optics is often understood as a branch of electromagnetism and its fundamental laws are derived from Maxwell's equations[4]. An electromagnetic field is characterized by an electric field **E** and a magnetic induction **B**. These fields are linked together by means of Maxwell's equations. In free space and in the absence of sources, each Cartesian component of **E** and **B** satisfies equation (1.1) independently[5]. The waves involved are known as electromagnetic waves.

Maxwell used his theory to deduce the speed of propagation of electromagnetic waves[6]. This speed was found to coincide with the experimentally determined speed of light. The extremely successful electromagnetic theory of light was born. The electromagnetic wave function is known as complex amplitude.

Wave-particle duality cannot be understood within the framework of the electromagnetic theory of light. The electromagnetic theory is unable to treat photons and the use of electromagnetic concepts can be inappropriate and misleading. Quantum physics is needed.

Quantum amplitude (or probability amplitude) is the term used to denote the wave function in quantum physics. A quantum amplitude, when multiplied by its complex conjugate, defines a relative probability density. The relative probability density describes the probability that a photon can interact within an infinitesimal area surrounding a particular point. The concepts of quantum amplitude and relative probability density are used in quantum physics to accommodate the existence of wave-particle duality.

## IMAGE FORMATION APPARATUS

During the process of real image formation, light wave components propagate away from an object, through an imaging system and ultimately form an image. Image forming light necessarily begins its journey by propagating away from the object plane.

A generalized optical real image formation apparatus is illustrated in Figure 1. The image formation apparatus is an ordinary diffraction-limited linear space-invariant imaging system.

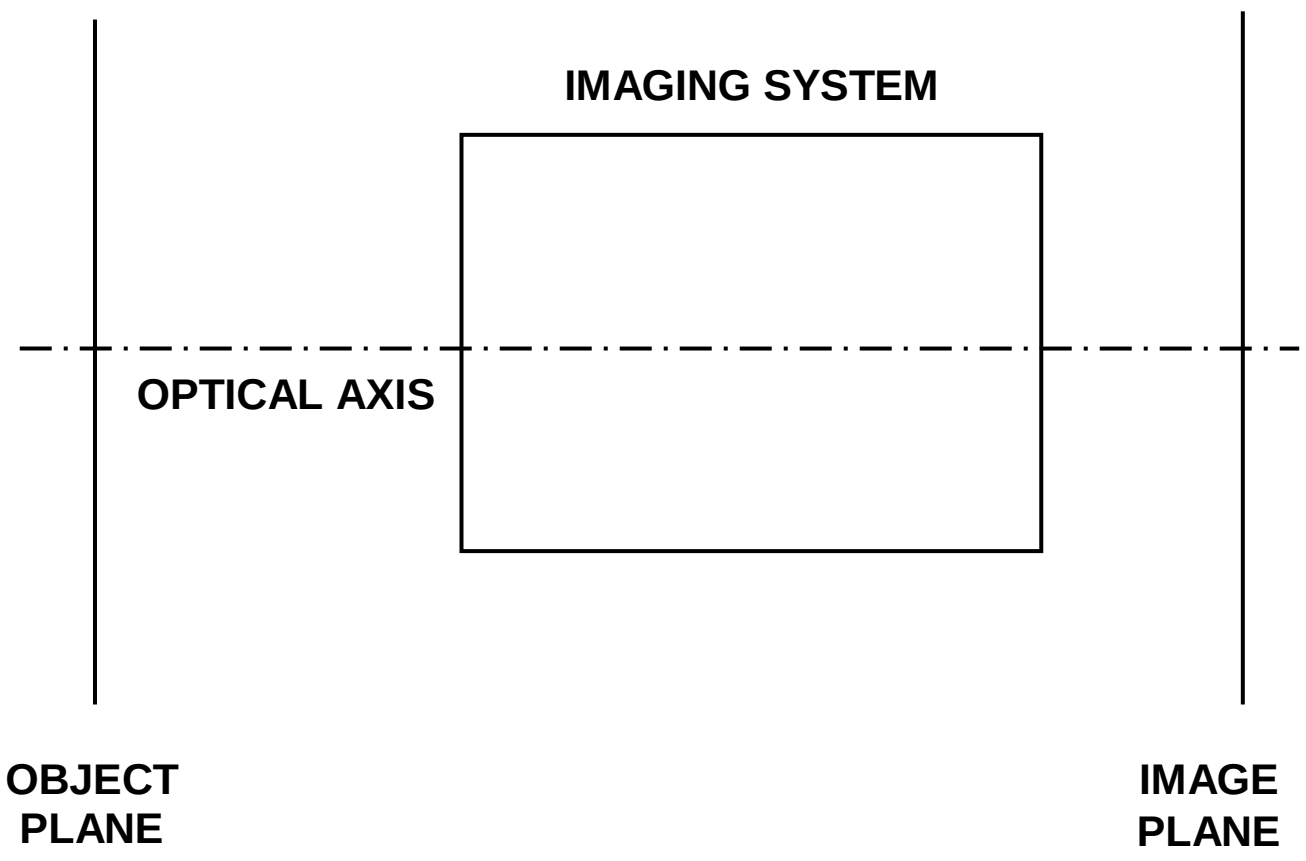

Figure 1. Generalized optical real image formation apparatus.

Initially, light that is incident upon the object plane is transmitted through or reflected from physical features in the object plane. Such light is distributed in a definite configuration on the side of the object plane nearest to the imaging system. A portion of this light propagates from the object plane to the imaging system. Some of the light that arrives at the imaging system propagates through it. Subsequently, the light that propagates through the imaging system propagates from the imaging system to the image plane. The resulting configuration of light on the image plane closely approximates a magnified (enlarged, unchanged or reduced and perhaps inverted) version of the configuration of light on the object plane.

## OVERVIEW OF OPTICAL REAL IMAGE FORMATION

As a result of interacting with physical features in the object plane, a definite configuration of light exists on the side of the object plane nearest to the imaging system. This configuration of light consists of photons that are transmitted through or reflected from the object plane at the instant they leave the object plane.

A configuration of light can exist whether or not it is observed. A real image may or may not be observed.

A right-handed rectangular coordinate system can be conveniently introduced to facilitate treating real image formation mathematically. In this coordinate system, let the object plane be the $x_o y_o$-plane and let the image plane be the $x_i y_i$-plane. The subscripts *o* and *i* designate the object side and the image side, respectively, of the imaging system.

Let the $z$-axis coincide with the optical axis defined by the imaging system. Let the positive $z$-direction be the direction from the object plane toward the imaging system. Finally, let the $z$-axis intersect the $x_o y_o$-plane at the origin of coordinates.

Let $\psi_\lambda(x, y; z_o)$ be the instantaneous quantum amplitude of the photons linked with light of wavelength $\lambda$ on the object plane. The relative probability density, given by

$$P_\lambda(x, y; z_o) = \psi_\lambda{}^*(x, y; z_o)\psi_\lambda(x, y; z_o) \tag{1.2}$$

where the superscript asterisk denotes complex conjugation, is linked to $\psi_\lambda(x, y; z_o)$. The relative probability density specifies the probability that a photon of wavelength $\lambda$ can interact within an infinitesimal area surrounding the point $(x, y; z_o)$ on the object plane. Each quantum amplitude component can be characterized as being monochromatic and coherent.

Some of the photons in the configuration of light on the object plane propagate away from the object plane and pass through the imaging system illustrated in Figure 1. In turn, some of these photons ultimately form a real image on the image plane.

Let $\sigma_\lambda(x, y; z_o)$ be the coherent impulse response of the real image formation apparatus for light of wavelength $\lambda$. Furthermore, let $\psi_\lambda(x, y; z_i)$ be the quantum amplitude of the photons associated with light of wavelength $\lambda$ on the image plane. This quantum amplitude is given by

$$\psi_\lambda(x, y; z_i) = \left(\frac{1}{m}\right)^2 \psi_\lambda\left(\frac{x}{m}, \frac{y}{m}; z_o\right) \otimes \sigma_\lambda\left(\frac{x}{m}, \frac{y}{m}; z_o\right) \tag{1.3}$$

where $\otimes$ denotes the convolution operator and $m$ denotes the lateral magnification of the image relative to the object; $m$ may be either negative or positive, depending upon whether the real image is inverted or not inverted relative to the object.

The relative probability density

$$P_\lambda(x, y; z_i) = \psi_\lambda{}^*(x, y; z_i)\,\psi_\lambda(x, y; z_i) \tag{1.4}$$

is linked to $\psi_\lambda(x, y; z_i)$. This relative probability density specifies the probability that a photon of wavelength $\lambda$ can interact within an infinitesimal area surrounding the point $(x, y; z_i)$ on the image plane.

Images are formed by the apparently random arrival of one photon at a time[7,8]. Photons associated with an image that is formed with sufficiently feeble light show no evidence of wave-like behavior. Rather, such images are built up by individual photons that act independently.

In addition to its particle-like properties, light has wave-like properties. The wave-like properties define the relative probability density that describes the probability that a photon can interact at a particular location. Each photon is associated with the entire relative probability distribution. An image that is formed with sufficiently strong light shows no evidence of being built up by individual photons. Rather, the entire image appears to be formed as a single event.

A very large number of photons contribute when acceptable real image formation occurs. The inherent granularity associated with the object and the image configurations of light vanishes. Consequently the density of photons that contribute to real image formation is proportional to the probability density.

Many authors provide treatments concerning the derivation of equation (1.3); see, for example, Collier, Burkhardt and Lin[9], Gaskill[10], Cathey[11] or Goodman[12]. In addition, equation (1.3) is derived in this paper where it is presented as equation (2.23). Equation (1.3) is the coherent real image equation for light of wavelength $\lambda$ .

Derivations of equation (1.3) are ordinarily based upon the Huygens-Fresnel principle. Following the Huygens-Fresnel principle, it has been widely assumed that only one illuminated point in an object is needed to support image-forming light. However, as shown in this paper, real image formation is dependent upon a previously unknown cooperative phenomenon. This cooperative phenomenon is such that two separate illuminated object points are required to support image-forming light.

The Huygens-Fresnel principle is not a law of physics and is furthermore known to be problematic[13]. In addition, *Ad hoc* approximations (such as initial approximations, the Fresnel approximations, the Fraunhofer approximation, etc.) are invoked in derivations of equation (1.3) that are based upon the Huygens-Fresnel principle. None of these derivations are based on fundamental physics.

A new theory of optical real image formation is developed in this paper. The Huygens-Fresnel principle and its attendant approximations are avoided in this development. Rather, the theory is based on fundamental quantum physics[14].

## II. IMAGE FORMATION

Equation (1.1) can be written, in the previously introduced three-dimensional rectangular coordinate system, as

$$\left(\frac{\partial^2}{\partial x^2}+\frac{\partial^2}{\partial y^2}+\frac{\partial^2}{\partial z^2}\right)\Psi_\lambda\left(x,y,z,t\right)=\frac{\partial^2}{\partial t^2}\Psi_\lambda\left(x,y,z,t\right) \tag{2.1}$$

In this equation, the Laplacian operator

$$\nabla^2=\frac{\partial^2}{\partial x^2}+\frac{\partial^2}{\partial y^2}+\frac{\partial^2}{\partial z^2} \tag{2.2}$$

has been expressed in rectangular coordinates.

## ANGULAR SPECTRUM OF PLANE WAVES

Two-dimensional Fourier transform pairs defined on arbitrary planes parallel to the object plane can be used[15] to treat the angular spectrum of plane waves. Accordingly, let

$$\tilde{\psi}_\lambda\left(\nu_x,\nu_y;z\right)=\int_{-\infty}^{+\infty}\int_{-\infty}^{+\infty}\psi_\lambda\left(x,y;z\right)\exp\left[-i2\pi\left(\nu_x x+\nu_y y\right)\right]dy\,dx \tag{2.3}$$

be the two-dimensional Fourier transform of $\psi_\lambda\left(x,y,z\right)$ on the plane defined by an arbitrary value of *z*;

$$\psi_\lambda\left(x,y;z\right)=\int_{-\infty}^{+\infty}\int_{-\infty}^{+\infty}\tilde{\psi}_\lambda\left(\nu_x,\nu_y;z\right)\exp\left[i2\pi\left(x\nu_x+y\nu_y\right)\right]d\nu_y\,d\nu_x \tag{2.4}$$

is the corresponding two-dimensional inverse Fourier transform. Here, the reciprocal variables $\nu_x$ and $\nu_y$, commonly known as spatial frequencies, are needed to define the two-dimensional Fourier transform pair.

The function $\tilde{\psi}_\lambda\left(\nu_x,\nu_y;z\right)$ is known as the angular spectrum of $\psi_\lambda\left(x,y;z\right)$. The value of $\tilde{\psi}_\lambda\left(\nu_x,\nu_y;z\right)$ on an arbitrary plane *z* in terms of its value on a parallel reference plane $z_p$ is given by

$$\tilde{\psi}_\lambda\left(\nu_x,\nu_y;z\right)=\tilde{\psi}_\lambda\left(\nu_x,\nu_y;z_p\right)\exp\left[i2\pi\left(z-z_p\right)\nu_z\right] \tag{2.5}$$

in accord with the theory of quantum diffraction[16]. Equation (2.5) describes diffraction from plane $z_p$ to plane $z$.

## PROPAGATION THROUGH THE IMAGE FORMATION APPARATUS

Refer to the optical real image formation apparatus illustrated in Figure 1. Propagation of wave components in the region between the object plane and the image plane and outside the imaging system can be treated in terms of the planes, linked to real image formation, illustrated in Figure 2. In Figure 2, the object and image planes are designated as the $z_o$- and $z_i$- planes, respectively. Furthermore, the entrance and exit planes of the imaging system are designated as the $z_{l-}$- and $z_{l+}$- planes, respectively. In addition, the wavelength $\lambda'$ of the light in the region between the imaging system and the image plane may differ from the wavelength $\lambda$ of light in the region between the object plane and the imaging system.

Spatial frequencies associated with the wave components when they enter the imaging system may differ from those associated with the same wave components when they exit the imaging system. Accordingly, let unprimed spatial frequencies be used in the region where $z_o < z < z_{l-}$ and let primed spatial frequencies be used in the region where $z_{l+} < z < z_i$. Treatment of the region where $z_o < z < z_{l-}$ can be separated from treatment of the region where $z_{l+} < z < z_i$.

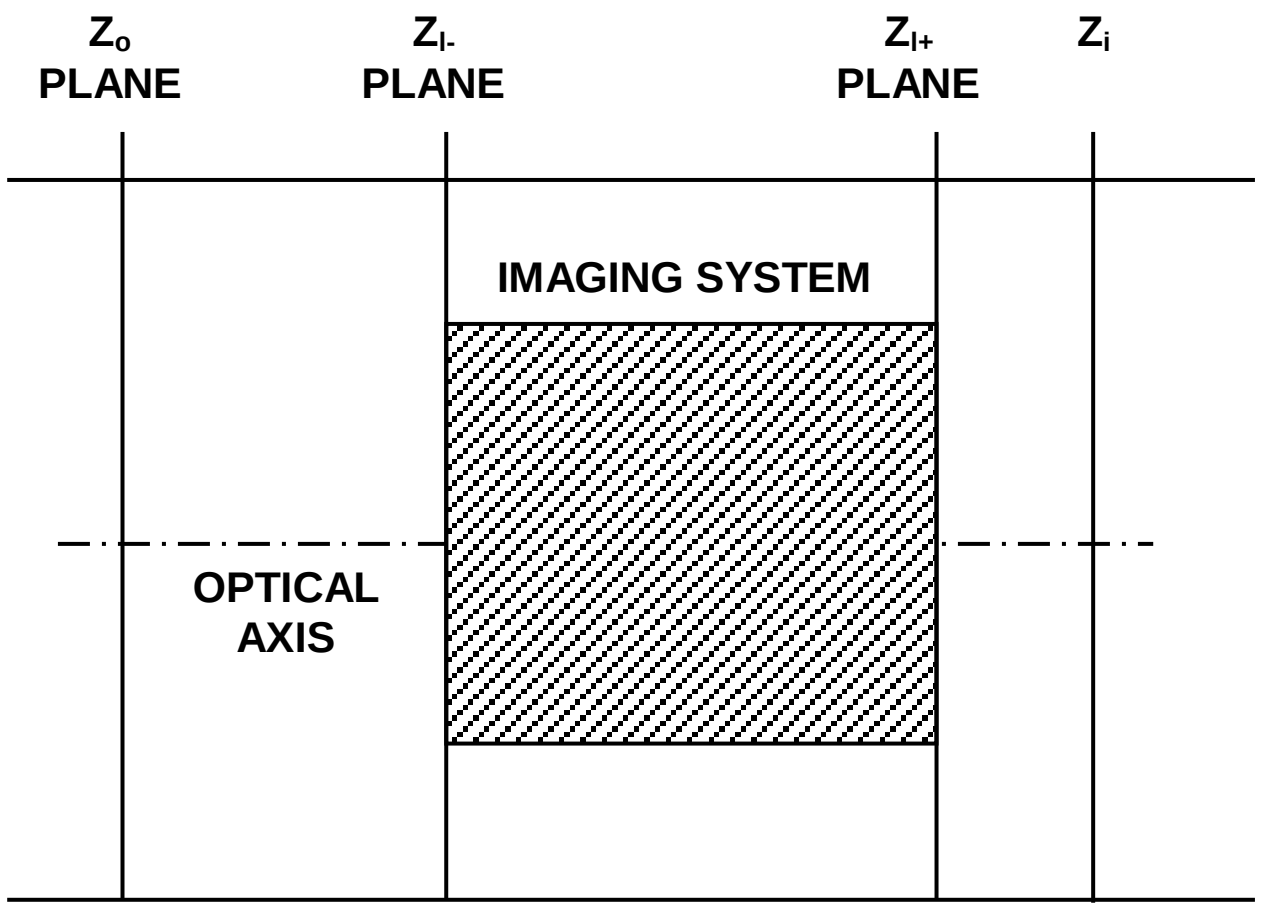

Figure 2. Image formation apparatus planes.

On the object side of the imaging system (in the region where $z_o < z < z_{l-}$), the wavelength of light used for image formation has been designated as $\lambda$. The possibility exists that this wavelength will be different on the image side of the imaging system due to a possible change in refractive index. To accommodate this possibility, the wavelength of light used will be designated as $\lambda'$ on the image side of the imaging system (in the region where $z_{l+} < z < z_i$).

The value of $\tilde{\psi}_\lambda\left(\nu_x,\nu_y;z\right)$ on an arbitrary plane $z$ in terms of its value on a parallel reference plane $z_p$ is given by equation (2.5) in the open interval $z_o < z < z_{l-}$ (inside the region where $z_o \le z \le z_{l-}$). Similarly, the value of $\tilde{\psi}_{\lambda'}\left(\nu'_x,\nu'_y;z\right)$ on an arbitrary plane $z$ in terms of its value on a parallel plane $z_{p'}$ is given by

$$\tilde{\psi}_{\lambda'}\left(\nu'_x,\nu'_y;z\right)=\tilde{\psi}_{\lambda'}\left(\nu'_x,\nu'_y;z_{p'}\right)\exp\left[i2\pi\left(z-z_{p'}\right)\nu'_z\right] \tag{2.6}$$

in the open interval $z_{l+} < z < z_i$ (inside the region where $z_{l+} < z < z_i$). The reference planes $z_p$ and $z_{p'}$ are determined independently of each other.

The relationship

$$\tilde{\psi}_\lambda\left(\nu_x,\nu_y;z_{l-}\right)=\tilde{\psi}_\lambda\left(\nu_x,\nu_y;z_o\right)\exp\left[i2\pi\left(z_{l-}-z_o\right)\nu_z\right] \tag{2.7}$$

follows when

$$\begin{pmatrix} z \\ z_p \end{pmatrix}=\begin{pmatrix} z_{l-} \\ z_o \end{pmatrix} \tag{2.8}$$

is substituted into equation (2.6). Similarly, the relationship

$$\tilde{\psi}_{\lambda'}\left(\nu'_x,\nu'_y;z_i\right)=\tilde{\psi}_{\lambda'}\left(\nu'_x,\nu'_y;z_{l+}\right)\exp\left[i2\pi\left(z_i-z_{l+}\right)\nu'_z\right] \tag{2.9}$$

follows when

$$\begin{pmatrix} z \\ z_{p'} \end{pmatrix} = \begin{pmatrix} z_i \\ z_{l+} \end{pmatrix} \tag{2.10}$$

is substituted into equation (2.6). After introducing

$$\begin{pmatrix} d_o \\ d_i \end{pmatrix} = \begin{pmatrix} z_{l-} - z_o \\ z_i - z_{l+} \end{pmatrix} \tag{2.11}$$

equations (2.7) and (2.9) can be written as

$$\tilde{\psi}_\lambda \left( \nu_x, \nu_y; z_{l-} \right) = \tilde{\psi}_\lambda \left( \nu_x, \nu_y; z_o \right) \exp\left( i 2\pi d_o \nu_z \right) \tag{2.12}$$

and

$$\tilde{\psi}_{\lambda'} \left( \nu'_x, \nu'_y; z_i \right) = \tilde{\psi}_{\lambda'} \left( \nu'_x, \nu'_y; z_{l+} \right) \exp\left( i 2\pi d_i \nu'_z \right) \tag{2.13}$$

respectively.

Let

$$\tilde{\sigma}_\lambda \left( \nu_x, \nu_y; z_o \right) = \int_{-\infty}^{+\infty} \int_{-\infty}^{+\infty} \sigma_\lambda \left( x, y; z_o \right) \exp\left[ -i 2\pi \left( \nu_x x + \nu_y y \right) \right] dy\, dx \tag{2.14}$$

be the coherent transfer function of the imaging system; $\sigma_\lambda \left( x, y; z_o \right)$ is the corresponding coherent impulse response introduced previously. The equation

$$\tilde{\psi}_\lambda \left( \nu'_x, \nu'_y; z_{l+} \right) = \tilde{\sigma}_\lambda \left( \nu_x, \nu_y; z_o \right) \tilde{\psi}_\lambda \left( \nu_x, \nu_y; z_{l-} \right) \tag{2.15}$$

relates the two-dimensional Fourier transform of the quantum amplitude on the exit plane of the imaging system to the two-dimensional Fourier transform of the quantum amplitude on the entrance plane of the imaging system. Simultaneous satisfaction of equations (2.12), (2.13) and (2.15) yields

$$\tilde{\psi}_\lambda \left( \nu'_x, \nu'_y; z_i \right) = \tilde{\sigma}_\lambda \left( \nu_x, \nu_y; z_o \right) \tilde{\psi}_\lambda \left( \nu_x, \nu_y; z_o \right) \exp\left[ i 2\pi \left( d_o \nu_z + d_i \nu'_z \right) \right] \tag{2.16}$$

readily.

Advanced imaging systems are designed such that (ideally)

$$d_o \nu_z + d_i \nu'_z = 0 \tag{2.17}$$

occurs. Substitution of equation (2.17) into equation (2.16) yields

$$\tilde{\psi}_\lambda\left(\nu'_x,\nu'_y;z_i\right)=\tilde{\sigma}_\lambda\left(\nu_x,\nu_y;z_o\right)\tilde{\psi}_\lambda\left(\nu_x,\nu_y;z_o\right) \tag{2.18}$$

readily. Equation (2.18) describes propagation of the angular spectrum from the object plane to the imaging system, through the imaging system, and then from the imaging system to the image plane.

### CHANGE OF SPATIAL FREQUENCIES

Spatial frequencies of the light that emerges from an imaging system may differ from the corresponding spatial frequencies of the light that enters the imaging system. Thus, spatial frequencies in the primed region, expressed in terms of those in the unprimed region, are given by

$$\begin{pmatrix}\nu_x'\\ \nu_y'\end{pmatrix}=\begin{pmatrix}\dfrac{\nu_x}{m}\\ \dfrac{\nu_y}{m}\end{pmatrix} \tag{2.19}$$

as a consequence of an optical disturbance's interaction with an ideal imaging system. Equation (2.18) reduces to

$$\tilde{\psi}_\lambda\left(\frac{\nu_x}{m},\frac{\nu_y}{m};z_i\right)=\tilde{\sigma}_\lambda\left(\nu_x,\nu_y;z_o\right)\tilde{\psi}_\lambda\left(\nu_x,\nu_y;z_o\right) \tag{2.20}$$

after the change of spatial frequencies has been invoked.

Many practical imaging systems have been designed to achieve very close approximations to the change of variables given by equation (2.19). Treatment of these imaging systems is beyond the scope of this paper.

### COHERENT REAL IMAGE EQUATION

The two-dimensional inverse Fourier transform of equation (2.20) is given by

$$\begin{aligned}&\int_{-\infty}^{+\infty}\int_{-\infty}^{+\infty}\tilde{\psi}_\lambda\left(\frac{\nu_x}{m},\frac{\nu_y}{m};z_i\right)\exp\left[i2\pi\left(x\nu_x+y\nu_y\right)\right]d\nu_y\,d\nu_x=\\&\qquad\int_{-\infty}^{+\infty}\int_{-\infty}^{+\infty}\tilde{\psi}_\lambda\left(\nu_x,\nu_y;z_o\right)\tilde{\sigma}_\lambda\left(\nu_x,\nu_y;z_o\right)\exp\left[i2\pi\left(x\nu_x+y\nu_y\right)\right]d\nu_y\,d\nu_x\end{aligned} \tag{2.21}$$

which yields

$$m^2\psi_\lambda\left(mx,my;z_i\right)=\psi_\lambda\left(x,y;z_o\right)\otimes\sigma_\lambda\left(x,y;z_o\right) \tag{2.22}$$

upon evaluation. Equation (2.22) can be written as

$$\psi_\lambda(x,y;z_i)=\left(\frac{1}{m}\right)^2\psi_\lambda\left(\frac{x}{m},\frac{y}{m};z_o\right)\otimes\sigma_\lambda\left(\frac{x}{m},\frac{y}{m};z_o\right) \tag{2.23}$$

which relates the quantum amplitude of the optical disturbance on the image plane to the quantum amplitude of the optical disturbance on the object plane. This result, the coherent real image equation, is identical with equation (1.3).

Equation (2.23) describes an optical real image as an ideal (undistorted) optical real image convolved with the impulse response linked to an imaging system. Except for special cases, the convolution operation leads to image distortion. Thus, features in the image are (ordinarily) approximately as wide as the sum of the widths of the two functions being convolved; fine features become indistinguishable (washed out). The distortion occurs as a consequence of the imaging system's impulse response and related transfer function.

## III. OBJECT PLANE QUANTUM AMPLITUDE

A definite configuration of light, designated as $\psi_\lambda(x,y;z_o)$, exists on the side of the object plane nearest to the imaging system. Every point in $\psi_\lambda(x,y;z_o)$ is specified relative to the origin of coordinates, i.e., by the intersection of the *z*-axis with the $x_o y_o$-plane.

Choose the *z*-axis such that it coincides with the imaging system’s optical axis (illustrated in Figure 1). This location is arbitrary relative to the location of $\psi_\lambda(x,y;z_o)$ and is unrelated to the object being imaged. Undesirable consequences of such arbitrariness can be dealt with by introducing the double delta function decomposition of $\psi_\lambda(x,y;z_o)$.

### DOUBLE DELTA FUNCTION DECOMPOSITION

Equivalent two-dimensional Dirac delta function decompositions of $\psi_\lambda(x,y;z_o)$ are given by

$$\psi_\lambda(x,y;z_o)=\int_{-\infty}^{+\infty}\int_{-\infty}^{+\infty}\psi_\lambda(\alpha,\beta;z_o)\,\delta(\alpha-x,\beta-y)\,d\beta\,d\alpha \tag{3.1}$$

and

$$\psi_\lambda(x,y;z_o)=\int_{-\infty}^{+\infty}\int_{-\infty}^{+\infty}\psi_\lambda(\eta,\xi;z_o)\,\delta(\eta-x,\xi-y)\,d\xi\,d\eta \tag{3.2}$$

where $(\alpha,\beta)$ and $(\eta,\xi)$ are arbitrary points within the quantum amplitude $\psi_\lambda(x,y;z_o)$. The angular spectrum of $\psi_\lambda(x,y;z_o)$, given by

$$\tilde{\psi}_\lambda\left(\nu_x,\nu_y;z_o\right)=$$

$$\left(\frac{1}{2}\right)\int_{-\infty}^{+\infty}\int_{-\infty}^{+\infty}\int_{-\infty}^{+\infty}\int_{-\infty}^{+\infty}\Bigg(\psi_\lambda\left(\alpha,\beta;z_o\right)\delta\left(\eta,\xi\right)\exp\left\{i2\pi\left[\left(\frac{\eta-\alpha}{2}\right)\nu_x+\left(\frac{\xi-\beta}{2}\right)\nu_y\right]\right\}$$
$$+\psi_\lambda\left(\eta,\xi;z_o\right)\delta\left(\alpha,\beta\right)\exp\left\{-i2\pi\left[\left(\frac{\eta-\alpha}{2}\right)\nu_x+\left(\frac{\xi-\beta}{2}\right)\nu_y\right]\right\}\Bigg) \tag{3.3}$$
$$X\;\exp\left\{-i2\pi\left[\left(\frac{\eta+\alpha}{2}\right)\nu_x+\left(\frac{\xi+\beta}{2}\right)\nu_y\right]\right\}d\beta\,d\alpha\,d\xi\,d\eta$$

can be obtained readily[17]. Each component in the integrand of equation (3.3) describes a two-dimensional complex exponential periodic function multiplied by a two-dimensional complex exponential phase factor.

The two-dimensional complex exponential phase factor in equation (3.3) exists because the intersection of the *z*-axis (the optical axis) and the $x_o y_o$-plane is arbitrary. The distance between the *z*-axis and the midpoint between the points $\left(\alpha,\beta\right)$ and $\left(\eta,\xi\right)$ is given by

$$M=\sqrt{\left(\frac{\eta+\alpha}{2}\right)^2+\left(\frac{\xi+\beta}{2}\right)^2} \tag{3.4}$$

As illustrated in Figure 3, the distance *M* is not a physical property of the distribution of light.

Each two-dimensional complex exponential periodic function in the integrand of equation (3.3) is periodic in two dimensions. The quantum amplitude spatial period is given by

$$T=\sqrt{\left(\frac{\eta-\alpha}{2}\right)^2+\left(\frac{\xi-\beta}{2}\right)^2} \tag{3.5}$$

and represents a physical property of the distribution of light.

The distance that separates the points $\left(\alpha,\beta\right)$ and $\left(\eta,\xi\right)$ is given by

$$\Delta=\sqrt{\left(\eta-\alpha\right)^2+\left(\xi-\beta\right)^2} \tag{3.6}$$

while the spatial period

$$T=\frac{\Delta}{2} \tag{3.7}$$

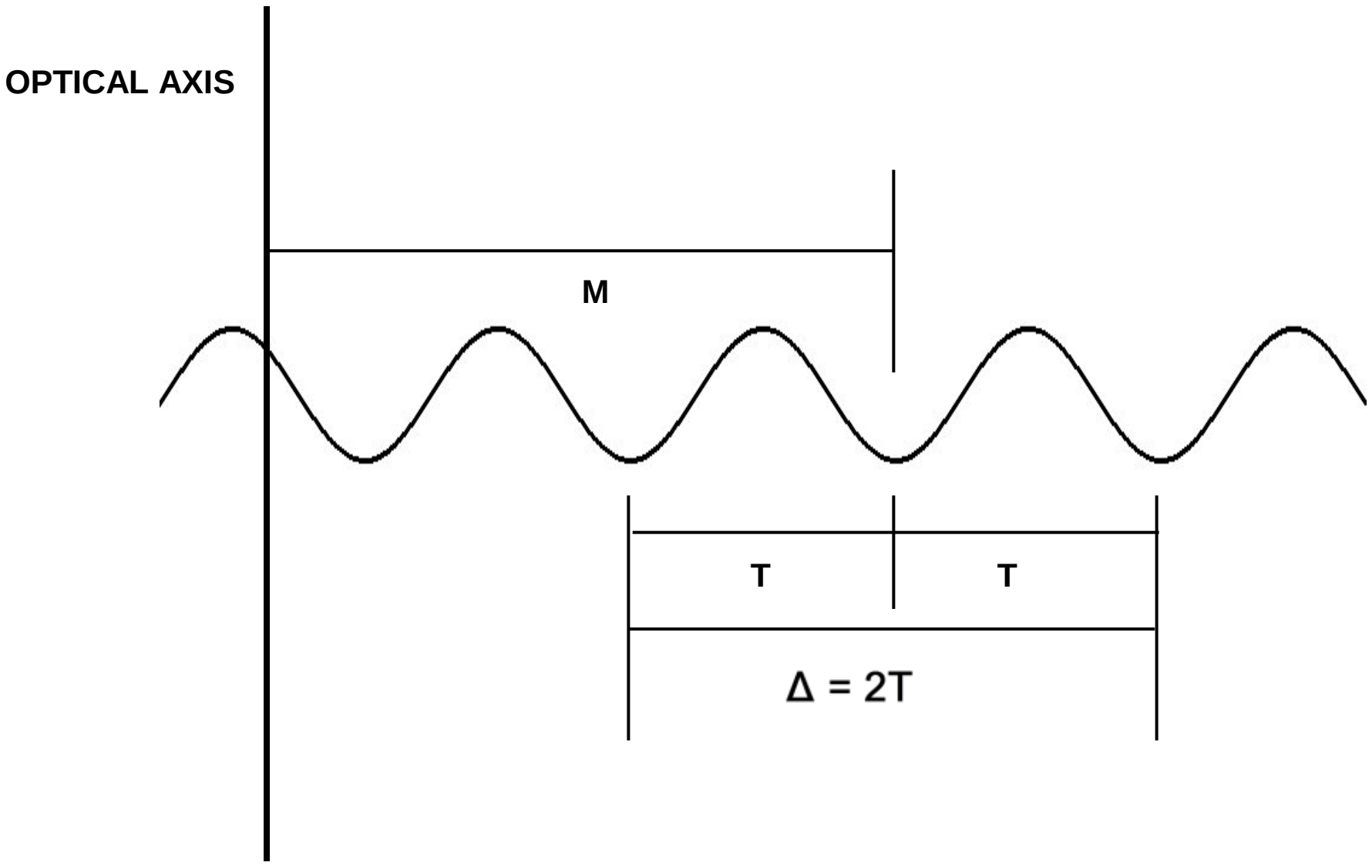

Figure 3. Object plane quantum amplitude component.

can be obtained by substituting equation (3.6) into equation (3.5). Thus, the quantum amplitude spatial period associated with two points in a distribution of light is equivalent to one-half the distance between the two points.

A single quantum amplitude component in the object plane is illustrated in Figure 3. Dimensions provided in this illustration enhance understanding $M$ , $T$ , $\Delta$ and, consequently, equation (3.3) as well.

**ILLUMINATION**

Consider the quantum amplitude associated with the two separated points $(\alpha,\beta)$ and $(\eta,\xi)$ in a configuration of light on the side of the object plane nearest to the imaging system. This quantum amplitude is given by

$$\psi_\lambda(x,y;z_o) = A_\lambda\delta(\alpha - x,\beta - y) + B_\lambda\delta(\eta - x,\xi - y) \tag{3.8}$$

where $A_\lambda$ and $B_\lambda$ are constants, for both points taken together. The individual quantum amplitudes are given by

$$\psi_\lambda(\alpha,\beta;z_o) = A_\lambda\delta(\alpha - x,\beta - y) \tag{3.9}$$

and

$$\psi_\lambda(\eta,\xi;z_o) = B_\lambda\delta(\eta - x,\xi - y) \tag{3.10}$$

for the two separated points. The two-dimensional Fourier transform of equation (3.8) can be shown to be given by[18]

$$\tilde{\psi}_{\lambda}\left(\nu_x,\nu_y;z_o\right)=\exp\left\{-i2\pi\left[\left(\frac{\eta+\alpha}{2}\right)\nu_x+\left(\frac{\xi+\beta}{2}\right)\nu_y\right]\right\}$$
$$X\left(A_{\lambda}\exp\left\{i2\pi\left[\left(\frac{\eta-\alpha}{2}\right)\nu_x+\left(\frac{\xi-\beta}{2}\right)\nu_y\right]\right\}\right. \qquad (3.11)$$
$$\left.+B_{\lambda}\exp\left\{-i2\pi\left[\left(\frac{\eta-\alpha}{2}\right)\nu_x+\left(\frac{\xi-\beta}{2}\right)\nu_y\right]\right\}\right)$$

Equation (3.11) can be written as

$$\tilde{\psi}_{\lambda}\left(\nu_x,\nu_y;z_o\right)=\left(A_{\lambda}+B_{\lambda}\right)\cos\left\{i2\pi\left[\left(\frac{\eta-\alpha}{2}\right)\nu_x+\left(\frac{\xi-\beta}{2}\right)\nu_y\right]\right\}$$
$$+i\left(A_{\lambda}-B_{\lambda}\right)\sin\left\{i2\pi\left[\left(\frac{\eta-\alpha}{2}\right)\nu_x+\left(\frac{\xi-\beta}{2}\right)\nu_y\right]\right\} \qquad (3.12)$$

by invoking Euler's formula.

The periodicity of $\tilde{\psi}_{\lambda}\left(\nu_x,\nu_y;z_o\right)$ is dependent upon $\tilde{\psi}_{\lambda}\left(\alpha,\beta;z_o\right)$ and $\tilde{\psi}_{\lambda}\left(\eta,\xi;z_o\right)$ both being nonzero. Consequently

$$\begin{pmatrix}A_{\lambda}\\B_{\lambda}\end{pmatrix}\neq\begin{pmatrix}0\\0\end{pmatrix} \qquad (3.13)$$

is required when $\tilde{\psi}_o\left(\nu_x,\nu_y;\lambda\right)$ is periodic. For the special case where

$$A_{\lambda}=B_{\lambda} \qquad (3.14)$$

equation (3.12) reduces to

$$\tilde{\psi}_{\lambda}\left(\nu_x,\nu_y;z_o\right)=2A_{\lambda}\cos\left\{i2\pi\left[\left(\frac{\eta-\alpha}{2}\right)\nu_x+\left(\frac{\xi-\beta}{2}\right)\nu_y\right]\right\} \quad \text{(Special Case)} \qquad (3.15)$$

easily.

Thoroughgoing investigation of light that is suitable for real image formation illumination is beyond the scope of this paper. Nevertheless, light that is either coherent or partially coherent where both illuminated points exist in the same area of coherence can be used satisfactorily.

#### INDIVIDUAL PHOTONS

Two illuminated object plane points are required, as a minimum, to form an optical real image. This is true either when the image is formed by a large number of photons in a single event or when the image is built up by individual photons. Evidently each individual photon is associated with illumination of both of the requisite object plane points. This occurs even in the extreme case where only one photon is involved and the distance between illuminated object points is arbitrarily large. Thus, each individual photon is nonlocal.

## IV. WAVE PROPAGATION AWAY FROM THE OBJECT PLANE

The three-dimensional Fourier transform of $\psi_\lambda(x,y,z)$

$$\tilde{\psi}_\lambda\left(\nu_x,\nu_y,\nu_z\right)=\int_{-\infty}^{+\infty}\int_{-\infty}^{+\infty}\int_{-\infty}^{+\infty}\psi_\lambda(x,y,z)\exp\left[-i2\pi\left(\nu_x x+\nu_y y+\nu_z z\right)\right]dz\,dy\,dx \tag{4.1}$$

can now be conveniently introduced. The corresponding three-dimensional inverse Fourier transform, given by

$$\psi_\lambda(x,y,z)=\int_{-\infty}^{+\infty}\int_{-\infty}^{+\infty}\int_{-\infty}^{+\infty}\tilde{\psi}_\lambda\left(\nu_x,\nu_y,\nu_z\right)\exp\left[i2\pi\left(x\nu_x+y\nu_y+z\nu_z\right)\right]d\nu_z\,d\nu_y\,d\nu_x \tag{4.2}$$

can also be conveniently introduced. The Fourier transform pair expresses $\tilde{\psi}_\lambda\left(\nu_x,\nu_y,\nu_z\right)$ and $\psi_\lambda(x,y,z)$ as linear combinations of three-dimensional complex exponential functions.

Equation and (4.2) describes the spatial dependence of a linear combination of plane waves that propagate in the direction associated with the direction cosines

$$\begin{pmatrix}\cos\theta_x\\ \cos\theta_y\\ \cos\theta_z\end{pmatrix}=\begin{pmatrix}\lambda\nu_x\\ \lambda\nu_y\\ \lambda\nu_z\end{pmatrix} \tag{4.3}$$

where $\theta_x$, $\theta_y$ and $\theta_z$ are the angles between the directions of wave propagation and the *x*-, *y*- and *z*-axes, respectively. The equation

$$\cos^2\theta_x+\cos^2\theta_y+\cos^2\theta_z=1 \tag{4.4}$$

constitutes a well-known fundamental property of direction cosines. Upon substitution of equation (4.3) into equation (4.4)

$$\lambda^2{\nu_x}^2+\lambda^2{\nu_y}^2+\lambda^2{\nu_z}^2=1 \tag{4.5}$$

follows readily.

Explicit expressions for the spatial frequencies $\nu_x$, $\nu_y$ and $\nu_z$ are provided by

$$\begin{pmatrix} \nu_x \\ \nu_y \\ \nu_z \end{pmatrix} = \begin{pmatrix} \nu \cos\theta_x \\ \nu \cos\theta_y \\ \nu \cos\theta_z \end{pmatrix} \tag{4.6}$$

Equation (4.6) is the column vector representation of the temporal frequency vector $\nu$.

As indicated earlier, the *z*-component of the direction of wave propagation is non-negative. Consequently,

$$\cos\theta_z \geq 0 \tag{4.7}$$

and, since $\lambda > 0$ (the wavelength of light is a positive entity),

$$\nu_z \geq 0 \tag{4.8}$$

follows when inequality (4.7) is substituted into the third row of equation (4.3)

After using some trigonometry,

$$\sin^2\theta_z = \cos^2\theta_x + \cos^2\theta_y \tag{4.9}$$

which is equivalent to

$$\sin\theta_z = \pm\sqrt{\cos^2\theta_x + \cos^2\theta_y} \tag{4.10}$$

follows from equation (4.4) readily.

Two equi-amplitude plane wave components are associated with each pair of illuminated points in the object plane. For each plane wave component that propagates away from the object plane at an angle $+\theta_z$ relative to the optical axis, an equi-amplitude plane wave component propagates at an angle $-\theta_z$ relative to the optical axis. The two equi-amplitude plane waves constitute a plane wave pair.

The propagation direction and a wavefront that is perpendicular to the propagation direction are illustrated in Figure 4 for an individual plane wave component. The wavefront is an infinite plane that is perpendicular to the plane of the figure and that extends out of the plane of the figure; only the trace of the wavefront on the plane of the figure is shown in the figure.

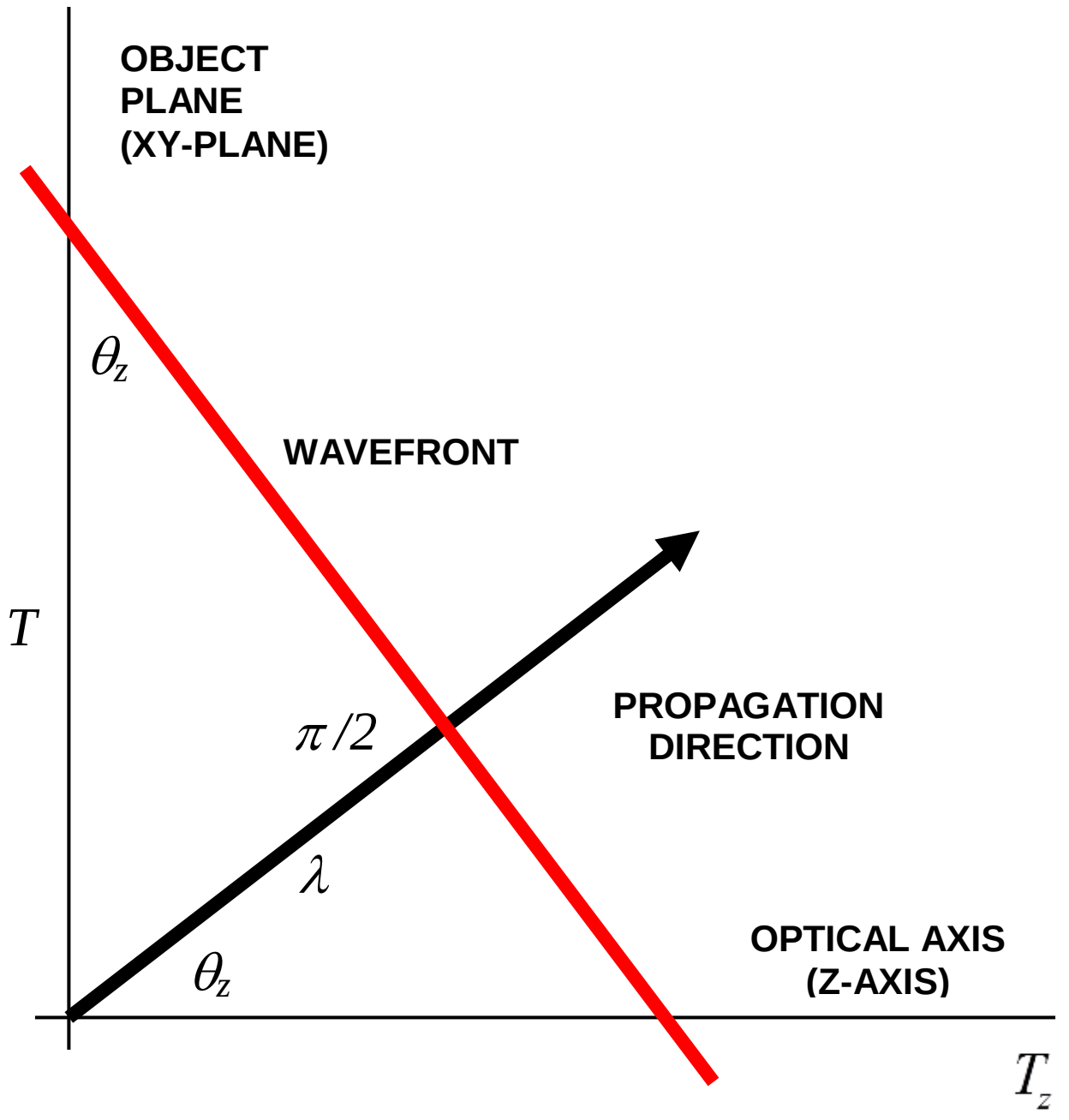

Figure 4. Plane wave propagation.

In Figure 4, the *z*-axis coincides with the optical axis defined by the imaging system. The object plane is rotated about the *z*-axis so that the line defined by the points $(\alpha, \beta)$ and $(\eta, \xi)$ is parallel to the plane of the figure. This line is associated with the spatial period *T*. The spatial period *T* is shown as the vertical axis in the figure and is also shown as the hypotenuse of a right triangle. The wavelength of light used, $\lambda$, is depicted in the figure as the distance between the origin and the wavefront introduced previously. In addition, $\theta_z$ is shown as the angle between the direction of wave propagation and the optical axis; $\lambda$, $T$ and $\theta_z$ have been introduced previously.

## V. MINIMUM SEPARATION REQUIREMENT

The relationship

$$\sin\theta_z = \frac{\lambda}{T} \tag{5.1}$$

can be established by inspecting Figure 4 and applying the definition of an angle's sine. Consequently,

$$\theta_z = \sin^{-1}\left(\frac{\lambda}{T}\right) \tag{5.2}$$

is the propagation angle associated with the spatial period *T*. Equation (5.2) can be written as

$$\theta_z = \sin^{-1}\left(\frac{2\lambda}{\Delta}\right) \quad (5.3)$$

where equation (3.7) has been invoked. The propagation angle (in degrees) is plotted in Figure 5 as a function of the separation distance (in wavelengths of the light used) between the points $(\alpha, \beta)$ and $(\eta, \xi)$.

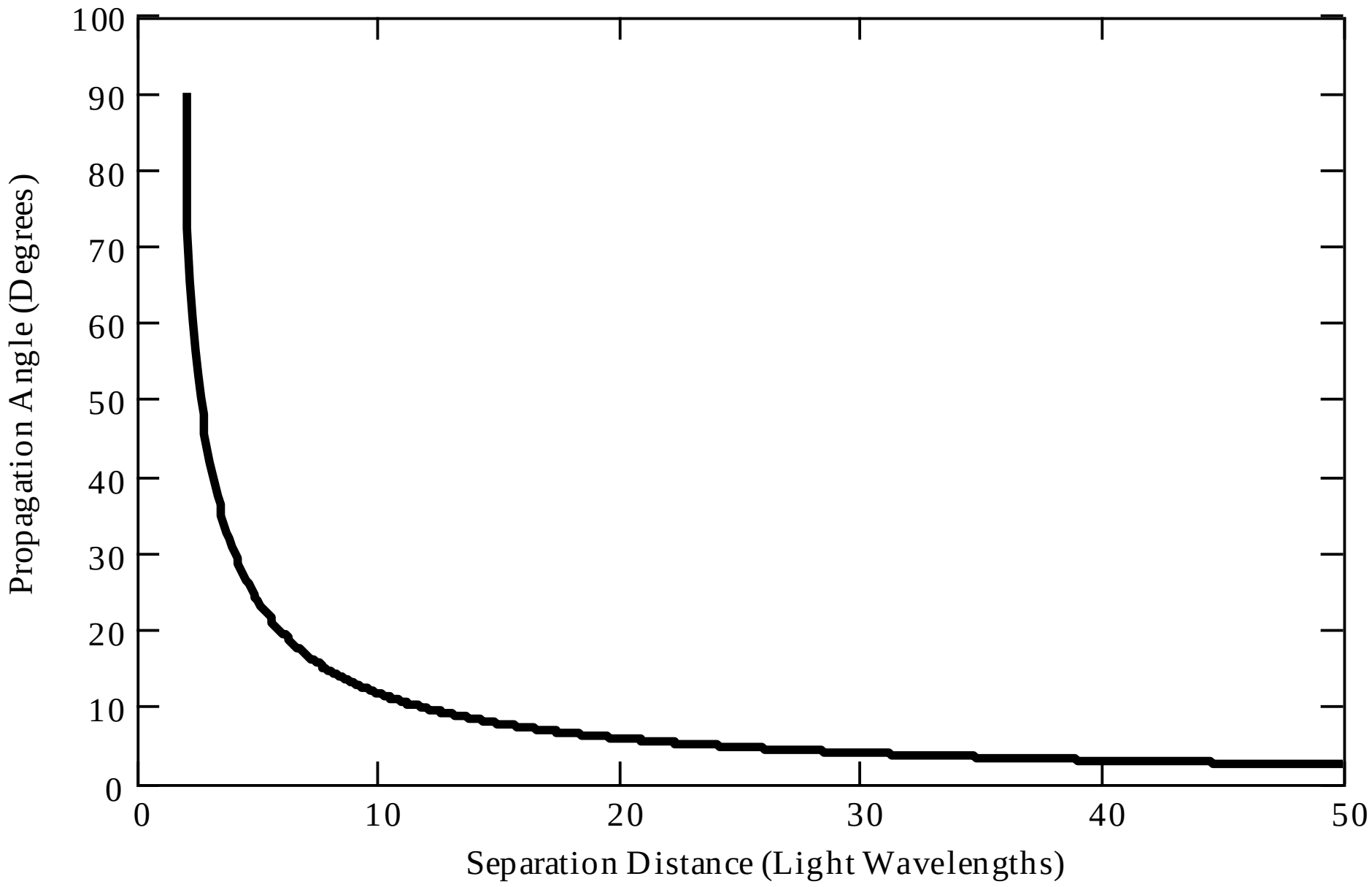

Figure 5. Propagation angle as a function of the separation distance between two points in the object.

Upon considering equation (5.3) or examining the curve in Figure 5, it can be determined readily that the minimum separation distance Δ required for light to propagate away from the object plane exceeds two wavelengths of the light used. Thus, satisfaction of the inequality

$$\Delta > 2\lambda \quad (5.4)$$

is required for light to propagate away from the object plane. As a consequence, propagation of light away from a single point (as hypothesized by the Huygens-Fresnel principle) does not occur. Two illuminated points in the object are required to form light waves that contribute to real image formation.

## VI. LATERAL MAGNIFICATION

Some properties of the light that emerges from an imaging system differ from those that enter the imaging system. As introduced previously, let unprimed parameters be used in the region between the object plane and the imaging system. Also, let primed parameters be used in the region between the imaging system and the image plane.

Let $n$ be the index of refraction in the unprimed region and let $n'$ be the index of refraction in the primed region. Furthermore, let $\lambda_S$ be the wavelength of the light used as observed in vacuum. Then

$$\begin{pmatrix} \lambda \\ \lambda' \end{pmatrix} = \begin{pmatrix} \dfrac{\lambda_S}{n} \\ \dfrac{\lambda_S}{n'} \end{pmatrix} \tag{6.1}$$

where $\lambda$ is the wavelength of the light used in the unprimed region and $\lambda'$ is the corresponding wavelength of light in the primed region.

Substitution of equation (6.1) into equation (5.1) leads to

$$\begin{pmatrix} \sin\theta_z \\ \sin\theta'_z \end{pmatrix} = \begin{pmatrix} \dfrac{\lambda_S}{nT} \\ \dfrac{\lambda_S}{n'T'} \end{pmatrix} \tag{6.2}$$

readily. In turn

$$\begin{pmatrix} \theta_z \\ \theta'_z \end{pmatrix} = \begin{pmatrix} \sin^{-1}\left(\dfrac{\lambda_S}{nT}\right) \\ \sin^{-1}\left(\dfrac{\lambda_S}{n'T'}\right) \end{pmatrix} \tag{6.3}$$

can be obtained from equation (6.2) quite easily. The angle between the direction of wave propagation and the *z*-axis is changed from $\theta_z$ to $\theta'_z$ as a result of the light passing through the imaging system.

The spatial period at the image plane of a plane wave that contributes to real image formation is given by

$$T' = mT \tag{6.4}$$

where the lateral magnification $m$ has been recalled. Substitution of equation (6.4) into the bottom row of equation (6.2) yields

$$\sin\theta'_z = \frac{\lambda_S}{n' m T} \tag{6.5}$$

which can be combined with the top row of equation (6.2) to obtain

$$m n' \sin\theta'_z = n \sin\theta_z \tag{6.6}$$

a result known as the optical invariant. The lateral magnification

$$m = \frac{n \sin\theta_z}{n' \sin\theta'_z} \tag{6.7}$$

can be obtained from equation (6.6) by simply rearranging terms.

## VII. COHERENT IMPULSE RESPONSE AND TRANSFER FUNCTION

The coherent impulse response associated with an arbitrary imaging system is given by

$$\sigma_\lambda(x, y; z_o) = \int_{-\infty}^{+\infty}\int_{-\infty}^{+\infty} \tilde{\sigma}_\lambda(\nu_x, \nu_y; z_o) \exp\left[i2\pi\left(x\nu_x + y\nu_y\right)\right] d\nu_y\, d\nu_x \tag{7.1}$$

for light of wavelength $\lambda$. Equation (7.1) is the two-dimensional inverse Fourier transform of the imaging system's coherent transfer function. The imaging system's coherent transfer function is given by

$$\tilde{\sigma}_\lambda(\nu_x, \nu_y; z_o) = \int_{-\infty}^{+\infty}\int_{-\infty}^{+\infty} \sigma_\lambda(x, y; z_o) \exp\left[-i2\pi\left(\nu_x x + \nu_y y\right)\right] dy\, dx \tag{7.2}$$

for light of wavelength $\lambda$. Equation (7.2) is the two-dimensional Fourier transform of the imaging system's coherent impulse response. Equations (7.1) and (7.2) constitute a two-dimensional Fourier transform pair.

Consider a diffraction-limited imaging system that has a clear aperture of arbitrary shape and size. For light of wavelength $\lambda$ this imaging system is associated with a passband and a coherent transfer function. The coherent transfer function is given by

$$\tilde{\sigma}_\lambda(\nu_x, \nu_y; z_o) = \begin{cases} 1 & \text{inside the passband} \\ 0 & \text{outside the passband} \end{cases} \tag{7.3}$$

Quantum amplitude components with spatial frequencies that are inside the passband of the imaging system pass through the imaging system and contribute to image formation. Quantum amplitude components with spatial frequencies that are outside the passband of the imaging system do not pass through the imaging system and do not contribute to image formation. Image distortion due to missing quantum amplitude components occurs.

Most imaging systems are circular. Although non-circular imaging systems exist, little would be gained by considering them in the present context. Henceforth, unless otherwise indicated, attention will be restricted to circular imaging systems.

Many authors provide treatments concerning coherent impulse responses and coherent transfer functions; see, for example, Gaskill[19] or Goodman[20]. A general treatment of coherent impulse responses and coherent transfer functions is beyond the scope of this paper.

## VIII. PASSBAND

Each plane wave component that arrives at the imaging system is associated with the spatial frequencies $\nu_x$, $\nu_y$ and $\nu_z$. These spatial frequencies are not independent. Thus

$$\nu_z = \sqrt{\frac{1}{\lambda^2} - \nu_x^{\,2} - \nu_y^{\,2}} \tag{8.1}$$

follows from equation (4.5) readily.

A maximum spatial frequency $\nu_c$ such that

$$\nu_z \leq \nu_c \tag{8.2}$$

is included in the passband that is linked to a circular imaging system. This spatial frequency is known as the spatial frequency cutoff of the imaging system. The inequality

$$\sqrt{\frac{1}{\lambda^2} - \nu_x^{\,2} - \nu_y^{\,2}} \leq \nu_c \tag{8.3}$$

is easily obtained by substituting equation (8.1) into inequality (8.2). The spatial frequency cutoff of an imaging system is the value of the spatial frequency beyond which the associated transfer function is zero.

Inside the imaging system's passband

$$0 \leq \sqrt{\frac{1}{\lambda^2} - \nu_x^{\,2} - \nu_y^{\,2}} \leq \nu_c \tag{8.4}$$

while

$$\nu_c < \sqrt{\frac{1}{\lambda^2} - \nu_x^2 - \nu_y^2} \tag{8.5}$$

outside the imaging system's passband. Accordingly

$$\tilde{\sigma}_\lambda\left(\nu_x, \nu_y; z_o\right) = \begin{cases} 1 & 0 \le \sqrt{\frac{1}{\lambda^2} - \nu_x^2 - \nu_y^2} \le \nu_c \\ 0 & \nu_c < \sqrt{\frac{1}{\lambda^2} - \nu_x^2 - \nu_y^2} \end{cases} \tag{8.6}$$

can be obtained by substituting equations (8.4) and (8.5) into equation (7.3).

Imaging systems with non-circular apertures are, in general, associated with multiple spatial frequency cutoffs. General treatment of such imaging systems is beyond the scope of this paper. The transfer functions associated with the imaging systems considered in this paper are very closely approximated by equation (8.6).

An imaging system that is associated with the spatial frequency cutoff $\nu_c$ is also associated with a spatial period cutoff $\tau_c$. The imaging system's spatial period cutoff, given by

$$\tau_c = \frac{1}{\nu_c} \tag{8.7}$$

is equal to the minimum spatial period of any quantum amplitude that can be transferred through the imaging system. This minimum spatial period is given by

$$T_c = \tau_c \tag{8.8}$$

Consequently, images of periodic quantum amplitude components with spatial periods $T$ such that

$$T \ge \tau_c \tag{8.9}$$

can be formed while images of quantum amplitude components with smaller spatial periods cannot be formed. All periodic quantum amplitude components with spatial periods that exceed $\tau_c$ pass through the imaging system and consequently contribute to real image formation.

Substitution of equation (3.7) into inequality (8.9) leads to

$$\Delta \ge 2\tau_c \tag{8.10}$$

readily. Accordingly, an image of two points in the object plane can be formed (the two points can be resolved) provided the distance between them equals or exceeds the minimum distance

$$\Delta_c = 2\tau_c \tag{8.11}$$

Two points in the object plane are, by definition, well-separated when inequality (8.10) is satisfied. Inequality (8.10) constitutes a fundamental two-point optical resolution criterion.

Circular imaging systems are endowed with a definite diameter. Accordingly, a circular imaging system restricts the propagation angle of the light that can enter it to a maximum allowed value $\theta_{zc}$. Similarly, a circular imaging system restricts the propagation angle of the light that can leave it to a maximum allowed value $\theta'_{zc}$. The cutoff propagation angles $\theta_{zc}$ and $\theta'_{zc}$ are the entrance angle (also known as the acceptance angle) and the exit angle, respectively, of the imaging system.

Equation (6.2) becomes

$$\begin{pmatrix} \sin\theta_{zc} \\ \sin\theta'_{zc} \end{pmatrix} = \begin{pmatrix} \dfrac{\lambda_S}{nT_c} \\ \dfrac{\lambda_S}{n'T'_c} \end{pmatrix} \tag{8.12}$$

when evaluated at cutoff. After rearranging equation (8.12) and substituting equation (8.8) into the result

$$\begin{pmatrix} \tau_c \\ \tau'_c \end{pmatrix} = \begin{pmatrix} \dfrac{\lambda_S}{n\sin\theta_{zc}} \\ \dfrac{\lambda_S}{n'\sin\theta'_{zc}} \end{pmatrix} \tag{8.13}$$

can be obtained.

The imaging system's object side numerical aperture $(\mathrm{NA})_o$ and image side numerical aperture $(\mathrm{NA})_i$, given by

$$\begin{pmatrix} (\mathrm{NA})_o \\ (\mathrm{NA})_i \end{pmatrix} = \begin{pmatrix} n\sin\theta_{zc} \\ n'\sin\theta'_{zc} \end{pmatrix} \tag{8.14}$$

can be conveniently introduced. Equation (6.7) becomes

$$m = \frac{n \sin \theta_{zc}}{n' \sin \theta'_{zc}} \tag{8.15}$$

when evaluated at cutoff. Substitution of equation (8.14) into equation (8.15) yields

$$m = \frac{(\mathrm{NA})_o}{(\mathrm{NA})_i} \tag{8.16}$$

easily.

Substitution of equation (8.14) into equation (8.13) leads to

$$\begin{pmatrix} \tau_c \\ \tau'_c \end{pmatrix} = \begin{pmatrix} \dfrac{\lambda_S}{(\mathrm{NA})_o} \\ \dfrac{\lambda_S}{(\mathrm{NA})_i} \end{pmatrix} \tag{8.17}$$

readily. In addition

$$\begin{pmatrix} \nu_c \\ \nu'_c \end{pmatrix} = \begin{pmatrix} \dfrac{(\mathrm{NA})_o}{\lambda_S} \\ \dfrac{(\mathrm{NA})_i}{\lambda_S} \end{pmatrix} \tag{8.18}$$

follows after substitution of equation (8.7) into equation (8.17). As shown by equations (8.17) and (8.18), knowledge of a circularly symmetric imaging system's object side and image side numerical apertures is sufficient to determine the imaging system's spatial period and spatial frequency cutoffs, respectively, for any particular wavelength of light.

An imaging system is often characterized in terms of its numerical aperture without distinguishing the object side numerical aperture from the image side numerical aperture. Rather, the term *numerical aperture* is used generically to treat either numerical aperture. The fundamental limits of performance for a circular imaging system are often expressed in terms of its numerical aperture without specifying the intended numerical aperture. Confusion can result.

## IX. RESOLUTION CRITERIA

A minimum separation distance between two distinguishable image features exists. Two image features that are sufficiently near each other merge together to form a single image feature. Two image features that can be distinguished as two image features are said to be resolved.

Comparison of the fundamental resolution criterion given by inequality (8.10) with two well-known classical resolution criteria can be made. These resolution criteria are the Abbė resolution criterion (ordinarily used for microscopes) and the Rayleigh resolution criterion (ordinarily used for telescopes and photolithography). The Rayleigh resolution criterion is often used to identify what is known as the classical diffraction limit of optical system performance. Neither of these criteria is based on fundamental physics; nevertheless, they are both roughly consistent with the fundamental resolution criterion introduced in this paper as inequality (8.10).

An overview of optical resolution has been published by A.J. den Decker and A. van den Bos[21]. These authors point out that optical resolution is not unambiguously defined and is interpreted in many ways. They go on to review the concept of optical resolution and many of its interpretations.

## FUNDAMENTAL RESOLUTION CRITERION

An image of two points in the object plane can be formed (the two points can be resolved) provided the distance between them equals or exceeds the cutoff distance. This fundamental two-point optical resolution criterion has been introduced previously by means of inequality (8.10).

Equation (8.12) can be rearranged so that

$$\begin{pmatrix} T_c \\ T'_c \end{pmatrix} = \begin{pmatrix} \dfrac{\lambda_S}{n\sin\theta_{zc}} \\ \dfrac{\lambda_S}{n'\sin\theta'_{zc}} \end{pmatrix} \tag{9.1}$$

results. Substitution of equation (8.14) into equation (9.1) yields

$$\begin{pmatrix} T_c \\ T'_c \end{pmatrix} = \begin{pmatrix} \dfrac{\lambda_S}{(\mathrm{NA})_o} \\ \dfrac{\lambda_S}{(\mathrm{NA})_i} \end{pmatrix} \tag{9.2}$$

readily. When expressed in terms of both unprimed and primed parameters, inequalities (8.9) and (8.10) lead to

$$\begin{pmatrix} T \\ T' \end{pmatrix} \geq \begin{pmatrix} \dfrac{\lambda_S}{(\mathrm{NA})_o} \\ \dfrac{\lambda_S}{(\mathrm{NA})_i} \end{pmatrix} \tag{9.3}$$

and

$$\begin{pmatrix} \Delta \\ \Delta' \end{pmatrix} \geq \begin{pmatrix} \dfrac{2\lambda_S}{(\mathrm{NA})_o} \\ \dfrac{2\lambda_S}{(\mathrm{NA})_i} \end{pmatrix} \tag{9.4}$$

respectively. Inequality (9.4) constitutes object side and image side fundamental two-point optical resolution criteria.

### ABBÉ RESOLUTION CRITERION

The Abbė resolution criterion can be written as

$$\begin{pmatrix} p \\ p' \end{pmatrix} \geq \begin{pmatrix} \dfrac{\lambda_S}{(\mathrm{NA})_o} \\ \dfrac{\lambda_S}{(\mathrm{NA})_i} \end{pmatrix} \tag{9.5}$$

and is applicable to a coherently illuminated periodic amplitude object of period $p$ as measured on the object plane and period $p'$ as measured on the image plane. Inequality (9.5) is mathematically equivalent to inequality (9.3) but it is fundamentally linked to spatial periods rather than two-point separations. An amplitude object where the spatial periods are physically identifiable is needed to apply the Abbė resolution criterion. By way of contrast, the fundamental resolution criterion is applicable to general objects.

### RAYLEIGH RESOLUTION CRITERION

The Rayleigh resolution criterion, which can be written as

$$\begin{pmatrix} \delta_R \\ \delta_R' \end{pmatrix} \geq \begin{pmatrix} \dfrac{1.22\lambda_S}{(\mathrm{NA})_o} \\ \dfrac{1.22\lambda_S}{(\mathrm{NA})_i} \end{pmatrix} \tag{9.6}$$

is a measure of the optical imaging system's performance. According to this criterion, an image of two points in the object plane can be formed (the two points can be resolved) provided the distance between them satisfies the Rayleigh criterion.

### CONVENTIONAL OPTICAL RESOLUTION

The conventional rationale that underlies the Rayleigh resolution criterion is presented in Appendix A. According to this rationale a typical image consists of many overlapping distributions of light. Each distribution of light is formed by the diffraction of coherent light from the imaging system's aperture. The image consists of the incoherent superposition of the coherently formed individual distributions of light. In accord with the Rayleigh criterion, two identical component distributions of light are barely resolved when they are separated by the distance $\delta_R$ (object side) or by the distance $\delta_R'$ (image side). The distance $\delta_R'$ is the laterally magnified version of the distance $\delta_R$ where the magnification is produced by the imaging system.

In accord with conventional real image formation theory, resolution and field depth can be understood (in disagreement with the nonlocal optical real image formation theory) in connection with an isolated three-dimensional distribution of light. Resolution is related to the width of the distribution of light in the focal plane while image depth is related to the length of the distribution of light perpendicular to the focal plane. As a consequence of this model, conventional resolution and field depth are closely linked to one another. According to the nonlocal optical real image formation theory, no such close linkage occurs.

## X. FIELD DEPTH

A configuration of light that closely approximates the configuration of light on the object (image) plane exists on various auxiliary planes that are parallel to and include the object (image) plane. The extreme range of distances over which these planes can be identified is the unobstructed object (image) field depth.

Unobstructed field depth occurs when the object (image) plane is transparent. Obstructed field depth occurs when the object (image) plane is opaque, and is ordinarily one-half the unobstructed field depth. Separate treatment of obstructed field depth is not included in this paper.

In the following analysis the object plane and the image plane are both treated as being transparent. The associated unobstructed field depth extends on both sides of the object (image) plane.

Table I. Parameters identified in Figure 6.

| | | |
|---|---|---|
| $\lambda'$ | — | Wavelength of light used (image side) |
| $\theta$ | — | Angle between the direction of propagation and a perpendicular to the image and auxiliary planes |
| $T'$ | — | Quantum amplitude spatial period of the sinusoidal component associated with the light wave (image side) |
| $z$ | — | Distance between the auxiliary plane and the image plane |
| $s$ | — | Longitudinal displacement of a spatial period on the auxiliary plane |
| $x$ | — | Distance in the direction defined by $T'$ |

**IMAGE FIELD DEPTH**

Consider optical real image formation where the image plane is transparent. Each plane wave pair that contributes to real image formation propagates through the image plane. These plane wave pairs also propagate through various planes that are parallel to the image plane. The plane wave pairs overlap on each of these planes. As a result, sinusoidal components that closely approximate corresponding sinusoidal components on the image plane are formed on auxiliary planes that are parallel to and sufficiently near the image plane. These sinusoidal components exist on both sides of the image plane and are considered to be part of the real image.

Superposition of the two members of a plane wave pair that overlap on a plane parallel to the image plane (an auxiliary plane) and a distance $z$ from it can now be considered. Geometric relationships defined by one member of the plane wave pair that contributes to real image formation are illustrated in Figure 6. Parameters identified in the figure are described in table I.

Let $A/2$ be the wave amplitude of each member of the plane wave pair and let $T'$ be the quantum amplitude spatial period of each member of the plane wave pair. Then

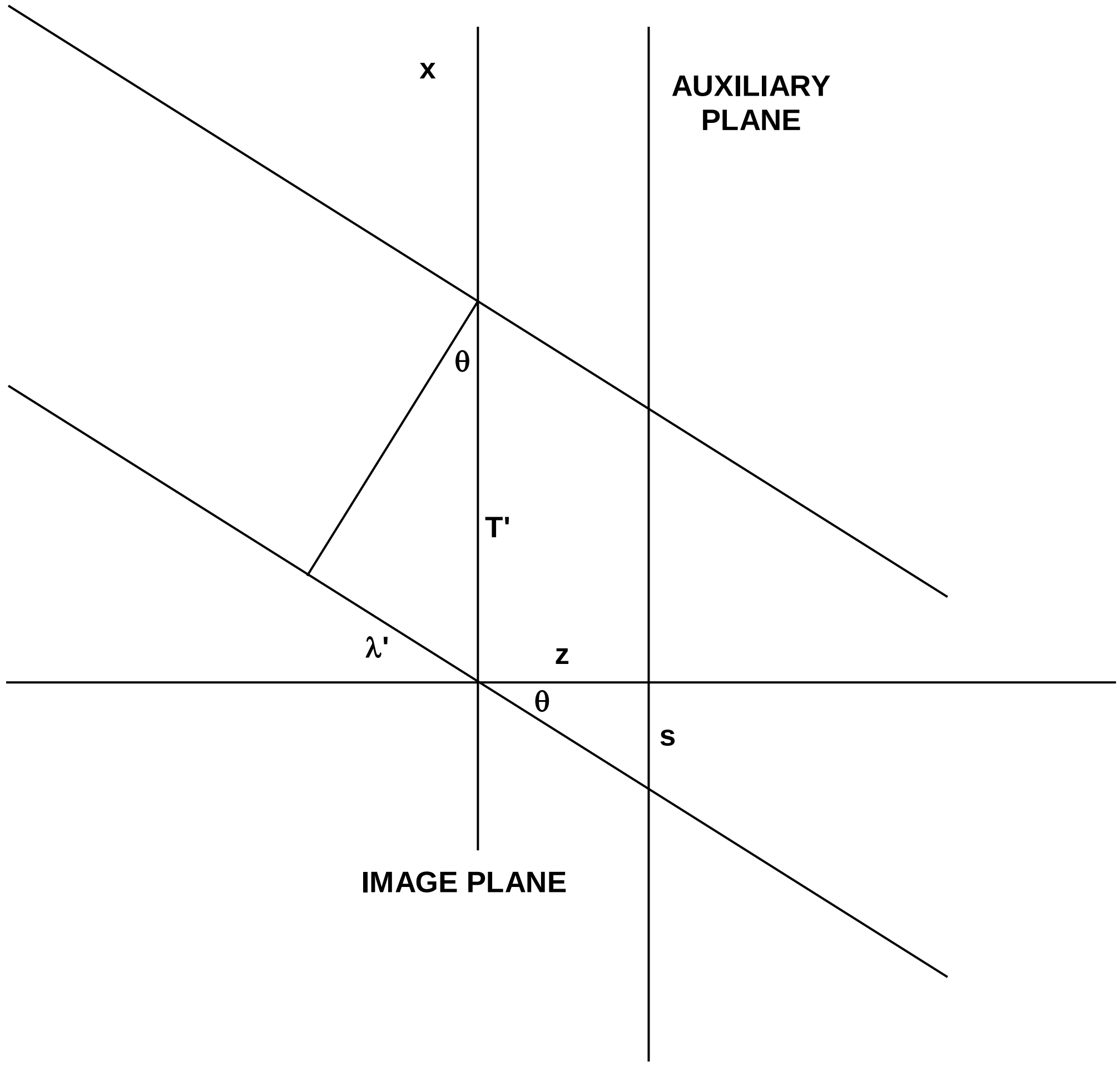

Figure 6. Plane wave propagation through the image plane and an auxiliary plane.

$$\psi_1 = \left(\frac{A}{2}\right)\cos\left[\left(\frac{2\pi}{T'}\right)(x+s)\right] \tag{10.1}$$

represents one member of the plane wave pair while

$$\psi_2 = \left(\frac{A}{2}\right)\cos\left[\left(\frac{2\pi}{T'}\right)(x-s)\right] \tag{10.2}$$

represents the other member of the plane wave pair. The superposition of both members of the plane wave pair can be described by

$$\psi = \psi_1 + \psi_2 \tag{10.3}$$

or

$$\psi = \left(\frac{A}{2}\right)\cos\left[\left(\frac{2\pi}{T'}\right)(x+s)\right] + \left(\frac{A}{2}\right)\cos\left[\left(\frac{2\pi}{T'}\right)(x-s)\right] \tag{10.4}$$

which reduces to

$$\psi = A\cos\left[\left(\frac{2\pi}{T'}\right)s\right]\cos\left[\left(\frac{2\pi}{T'}\right)x\right] \tag{10.5}$$

with little effort.

The relationships

$$s = z\tan\theta \tag{10.6}$$

$$\tan\theta = \frac{\sin\theta}{\cos\theta} \tag{10.7}$$

$$\sin\theta = \frac{\lambda'}{T'} \tag{10.8}$$

and

$$\cos\theta = \frac{\sqrt{T'^2 - \lambda'^2}}{T'} \tag{10.9}$$

can be obtained by inspecting Figure 6. Hence,

$$\tan\theta = \frac{\lambda'}{\sqrt{T'^2 - \lambda'^2}} \tag{10.10}$$

and

$$s = \left(\frac{\lambda'}{\sqrt{T'^2 - \lambda'^2}}\right) z \tag{10.11}$$

follow readily.

Substitution of equation (10.11) into equation (10.5) yields

$$\psi = A\cos\left[\left(\frac{\lambda'}{\sqrt{T'^2 - \lambda'^2}}\right)\left(\frac{2\pi}{T'}\right) z\right]\cos\left[\left(\frac{2\pi}{T'}\right) x\right] \tag{10.12}$$

easily. The first cosine function on the right hand side of equation (10.12) is spatially periodic where

$$\Omega = \frac{T'\sqrt{T'^2 - \lambda'^2}}{\lambda'} \tag{10.13}$$

is the function's spatial period. The range of values of $z$ for which changes in the value of $\psi$ are acceptable can be expressed as

$$z = \pm\frac{T'\sqrt{T'^2 - \lambda'^2}}{K\lambda'} \tag{10.14}$$

where $K$ is a subjectively determined constant.

The depth of image field can now be defined as

$$D' = z_+ - z_- \tag{10.15}$$

where

$$z_+ = +\frac{T'\sqrt{T'^2 - \lambda'^2}}{K\lambda'} \tag{10.16}$$

and

$$z_{-} = -\frac{T'\sqrt{T'^2 - \lambda'^2}}{K\lambda'} \qquad (10.17)$$

have been introduced. Substitution of equations (10.16) and (10.17) into equation (10.15) yields

$$D' = \frac{2T'\sqrt{T'^2 - \lambda'^2}}{K\lambda'} \qquad (10.18)$$

easily. The depth of image field is given by equation (10.18).

### OBJECT FIELD DEPTH

The analysis used to derive equation (10.18) for the depth of image field is also applicable to deriving analogous expressions for the depth of object field. Primed parameters are used for the image side analysis; unprimed counterparts are used for the object side analysis. Thus, the object side wavelength of light used is designated as $\lambda$ (unprimed) and the object side quantum amplitude spatial period is designated as $T$ (unprimed) in the treatment rather than their image side (primed) counterparts.

The object side counterpart of equation (10.18) is

$$D = \frac{2T\sqrt{T^2 - \lambda^2}}{K\lambda} \qquad (10.19)$$

The depth of object field is given by equation (10.19).

### VALUE OF K

As introduced previously, $K$ is a subjectively determined constant. Selecting a suitable value for $K$ is perhaps best done on the basis of experience. Possibly the value $K = 16$ may be suitable as a tentative useful value for $K$.

### TWO POINT SEPARATION

The quantum amplitude spatial period associated with two points in a configuration of light is equivalent to one-half the distance between the two points. This distance is given by equation (3.7) for points on the object plane. Consequently,

$$\begin{pmatrix} T \\ T' \end{pmatrix} = \begin{pmatrix} \dfrac{\Delta}{2} \\ \dfrac{\Delta'}{2} \end{pmatrix} \qquad (10.20)$$

when the object side and its image side (primed) counterpart are both considered. Substitution of equation (10.20) into equations (10.19) and (10.18) leads to

$$\begin{pmatrix} D \\ D' \end{pmatrix} = \begin{pmatrix} \dfrac{\Delta\sqrt{\Delta^2 - 4\lambda^2}}{2K\lambda} \\ \dfrac{\Delta'\sqrt{\Delta'^2 - 4\lambda'^2}}{2K\lambda'} \end{pmatrix} \tag{10.21}$$

The depth of object (image) field is independent of the imaging system.

### CONVENTIONAL IMAGE FIELD DEPTH

Equations that are commonly used to estimate the field depth (object and image) associated with real image formation are dependent upon the numerical aperture – designated as NA (without regard for the side of the imaging system involved) – of the imaging system that is used. Specifically

$$D_C = \frac{\lambda}{(\mathrm{NA})^2} \tag{10.22}$$

is the equation conventionally used to estimate the depth of object field associated with real image formation[22]. Similarly,

$$D'_C = \frac{\lambda'}{(\mathrm{NA})^2} \tag{10.23}$$

is the equation that is conventionally used to estimate the depth of image field associated with real image formation These equations are substantially different than the counterpart equations derived in this paper.

## XI. REFERENCE DISTRIBUTION REAL IMAGE FORMATION

Reference distribution real image formation (RIF) is achieved by using a reference distribution of light in addition to the usual subject distribution. This assures that two well-separated illuminated points are used to form each light wave component that propagates away from the object plane and ultimately contributes to real image formation. One point lies within the subject distribution and the other point lies within the reference distribution. The distance between these points is designed to assure that it exceeds the minimum needed for real image formation. Thus, inequality (8.10) is satisfied intentionally for every pair of points such that one point lies within the subject distribution and the other point lies within the reference distribution. An image of both distributions of light is formed.

The spatial frequencies of the component light waves that contribute to RIF lie within a well-defined finite bandwidth. The imaging system is designed such that it transfers the wave components with spatial frequencies that lie within this bandwidth to the image plane. These wave components are superposed on the image plane to form a real image.

#### RIF REAL IMAGE EQUATION

All spatial frequencies linked to the combined reference and subject distributions are included in $\tilde{\psi}_\lambda\left(\nu_x,\nu_y;z_o\right)$ when RIF is used. The imaging system is chosen to assure that these spatial frequencies lie within its passband. Thus

$$\tilde{\psi}_\lambda\left(\nu_x,\nu_y;z_o\right)\tilde{\sigma}_\lambda\left(\nu_x,\nu_y;z_o\right)=\tilde{\psi}_\lambda\left(\nu_x,\nu_y;z_o\right) \tag{11.1}$$

for reference distribution image formation. Consequently, equation (2.20) reduces to

$$\tilde{\psi}_\lambda\left(\frac{\nu_x}{m},\frac{\nu_y}{m};z_i\right)=\tilde{\psi}_\lambda\left(\nu_x,\nu_y;z_o\right) \tag{11.2}$$

and equation (2.23) reduces to

$$\psi_\lambda\left(x,y;z_i\right)=\left(\frac{1}{m}\right)^2\psi_\lambda\left(\frac{x}{m},\frac{y}{m};z_o\right) \tag{11.3}$$

for RIF. Equation (11.3) is the RIF real image equation for light of wavelength $\lambda$. No convolution operation is associated with the RIF real image equation.

The RIF real image equation defines a magnified (enlarged, unchanged or reduced and perhaps inverted) real image of both the subject distribution and the reference distribution. Magnified versions of all spatially periodic components linked to one point in the object distribution and another point in the reference distribution are included in the real image. No real image distortion due to missing spatially periodic components occurs. Resolution is complete.

#### OBJECT PLANE CONFIGURATION

A configuration of light that is suitable for achieving RIF is illustrated in Figure 7. The configuration of light exists on the side of the object plane nearest to the imaging system and is confined to the interiors of the circles shown in the figure. Light throughout the configuration is coherent or partially coherent. All points in both the subject distribution and the reference distribution are in the same area of coherence.

As indicated in the figure, the subject distribution exists inside a circular area of diameter $D_S$ while the reference distribution exists inside a circular area of diameter $D_R$. As introduced previously, let $\lambda_s$ be the wavelength of light used, as measured in vacuum, and let $(NA)_o$ be the object side numerical aperture of the imaging system. Ideal RIF occurs when the criteria

$$S > \frac{2\lambda_S}{(\mathrm{NA})_o} \tag{11.4}$$

$$D_S \leq \frac{2\lambda_s}{(\mathrm{NA})_o} \tag{11.5}$$

and

$$D_R \leq \frac{2\lambda_s}{(\mathrm{NA})_o} \tag{11.6}$$

are realized physically. Noisy RIF occurs when either $D_S$ or $D_R$ or both $D_S$ and $D_R$ do not satisfy the foregoing inequalities.

The fundamental resolution criterion, given by equation (9.4), is satisfied when inequality (11.4) is satisfied; an image of every object point and every image point is formed. The fundamental

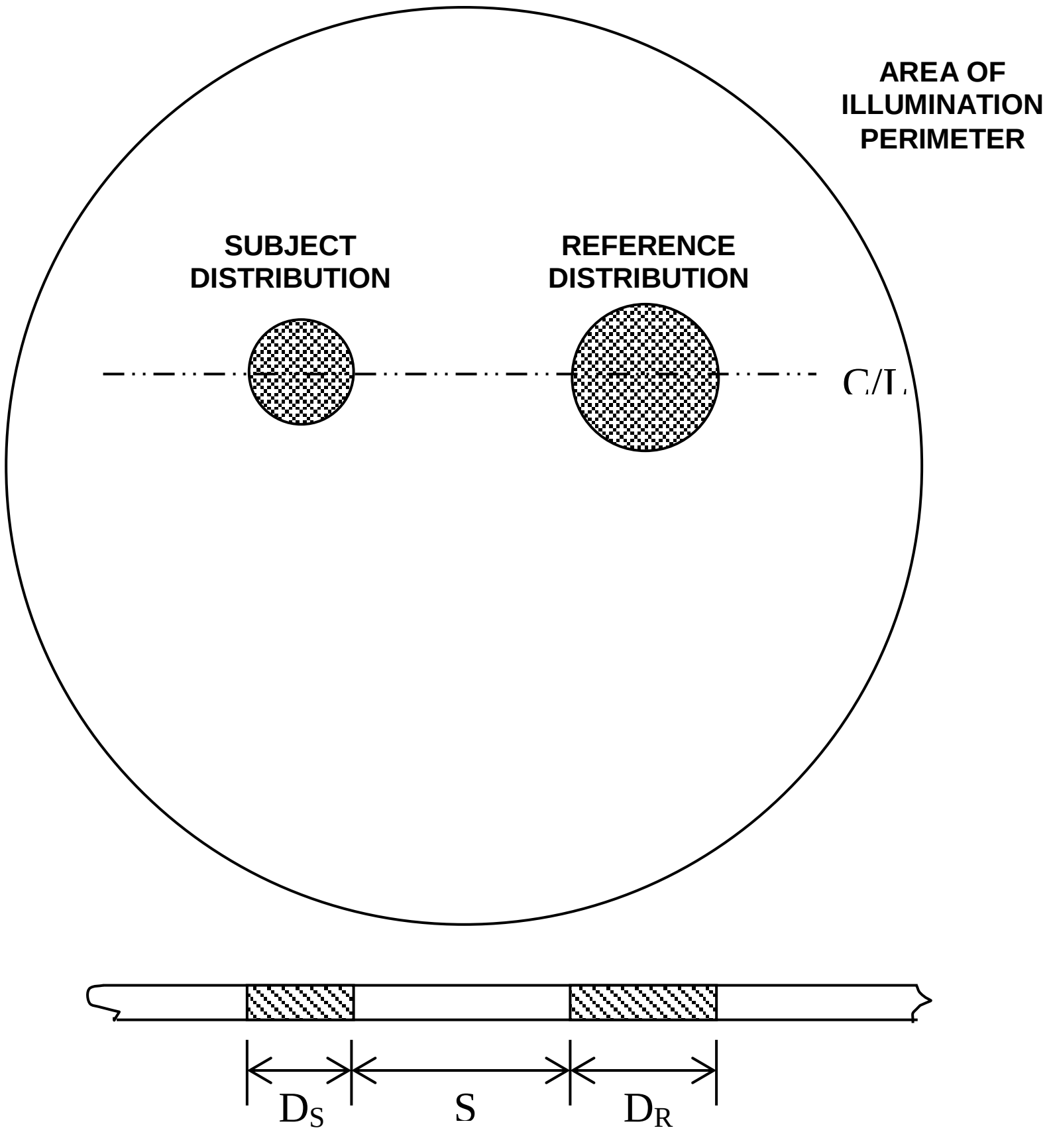

Figure 7. RIF object plane configuration: plan and front elevation views.

resolution criterion is not satisfied when inequalities (11.5) and (11.6) are satisfied.

The RIF configuration that has been presented is by no means the only possible RIF configuration that can be devised. Treatment of other possible RIF configurations is beyond the scope of this paper.

**NOISY RIF CONCEPT**

Consider the conceptual illustration of transmission noisy RIF provided in Figure 8. Initially, light is incident (from the left) upon an opaque screen with two apertures in it. The side of the screen nearest to the imaging system (shown as a lens) serves as an object plane.

Coherent or partially coherent light passes through the apertures to form two distributions of light on the object plane. These distributions of light are, ideally, the same size and shape as the apertures. The distributions of light are designated as the subject distribution and the reference distribution.

Taken separately, the propagation angles of the light that travels away from either of the two distributions of light are larger than the acceptance angle of the imaging system. Such light does not pass through the imaging system and consequently does not contribute to image formation.

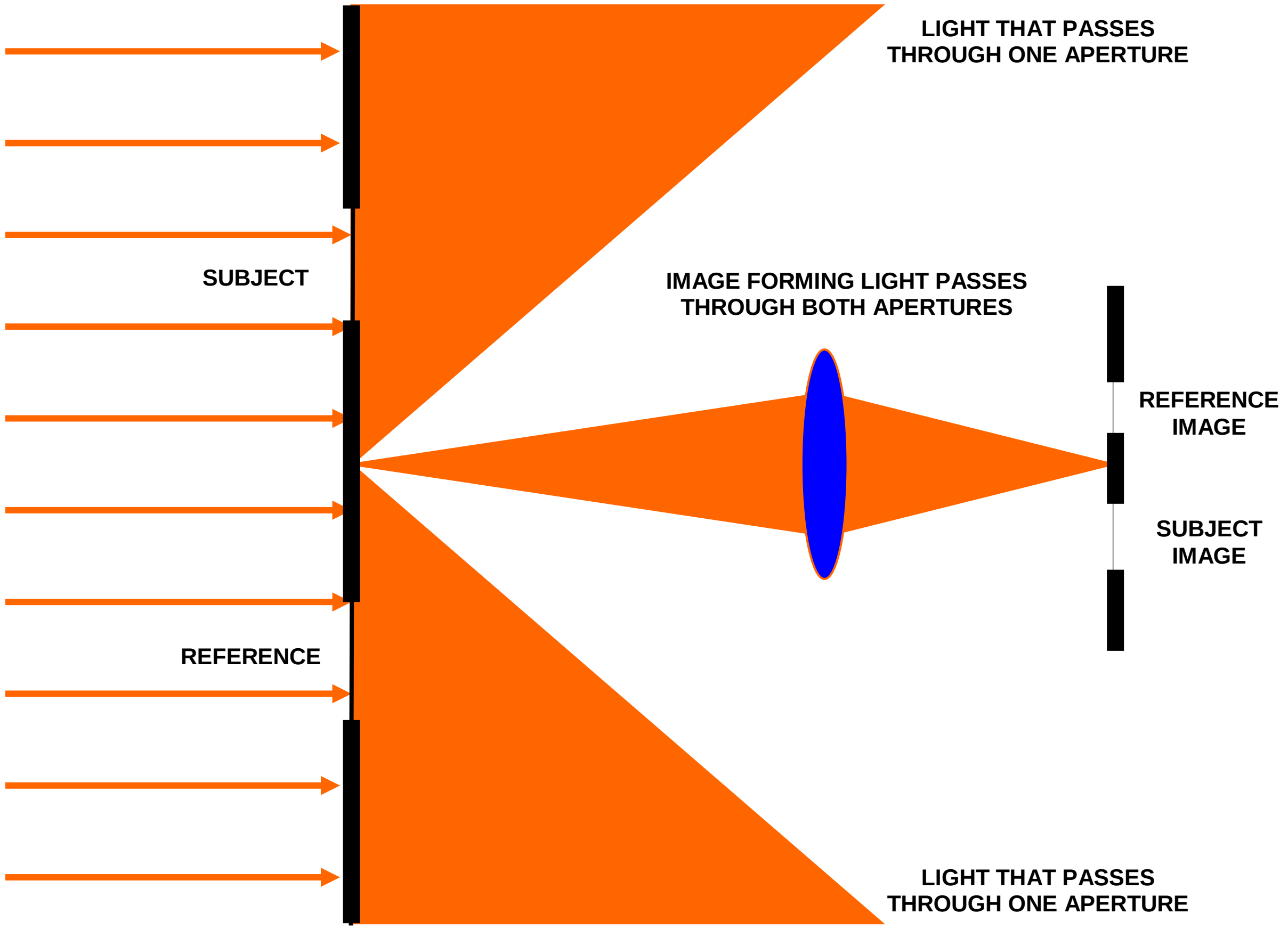

Figure 8. Conceptual noisy RIF process.

No two points in either the subject distribution or the reference distribution are sufficiently separated to contribute to image formation. Every pair of points such that one point is in the subject distribution and one point is in the reference distribution is sufficiently separated to contribute to image formation.

Taken together, the propagation angles of the light that travels away from the combined distributions of light are smaller than the acceptance angle of the imaging system. This light defines a finite bandwidth and is transferred, without amplitude or phase distortion, through the imaging system. Consequently, light that propagates away from the combined distributions of light contributes to undistorted real image formation.

## XII. CONCLUSION

A nonlocal theory of optical real image formation has been developed from the fundamental physics associated with an optical real image formation apparatus. The theory shows how two separated illuminated object points are fundamentally required for real image formation. This finding is in distinct contrast to the ordinarily accepted assumption that only one object point is involved. In accord with the new theory, optical real images with a very large field depth (object and image) and no fundamental resolution limit can be formed.

Each quantum amplitude component that contributes to real image formation is associated with a pair of illuminated points in the object. Both illuminated points are inside the area of coherence associated with the quantum amplitude component. Image-forming light is either coherent or partially coherent. Single photons are capable of providing image-forming light.

A fundamental two-point resolution criterion was derived as part of the new theory and compared with the Abbė and Rayleigh resolution criteria. Although neither of these criteria is based on fundamental physics, they are both roughly consistent with the fundamental resolution criterion introduced in this paper.

Field depths for both the object and image were shown to be independent of the imaging system used and to be substantially different than they are traditionally thought to be.

Reference distribution real image formation (RIF) has been introduced. RIF is achieved by using a reference distribution of light in addition to the usual subject distribution. This assures that two well-separated illuminated points are used to form each light wave component that propagates away from the object plane and ultimately contributes to real image formation. One point lies within the subject distribution and the other point lies within the reference distribution.

The spatial frequencies of the component light waves that contribute to RIF lie within a well-defined finite bandwidth. The imaging system transfers the wave components with spatial frequencies that lie within this bandwidth to the image plane. These wave components are superposed on the image plane to form a real image.

Real images formed by means of RIF are described by the RIF real image equation. The RIF real image equation defines a magnified (enlarged, unchanged or reduced and perhaps inverted) real image of both the subject distribution and the reference distribution. No convolution operation is involved. Magnified versions of all spatially periodic components included in the object are included in the real image. No real image distortion due to missing spatially periodic components occurs. Resolution is complete when real images are formed by means of RIF.

# APPENDIX A: CONVENTIONAL OPTICAL REAL IMAGE RESOLUTION

### INTRODUCTION

A brief review of the generally accepted treatment of resolution associated with optical real image formation (in disagreement with the nonlocal optical real image formation theory) is provided in this appendix. Thus, resolution associated with a spherical wave that converges to a geometric focus after passing through a circular aperture, as conventionally understood, is reviewed. Such a wave is often created by a focusing element, where the edge of the focusing element constitutes the edge of the aperture. For this conventional treatment, the angle at the geometric focus subtended by the circular aperture is restricted to being very small.

Referring to Figure A-1, a uniform spherical monochromatic light wave of wavelength $\lambda'$ converges toward a geometric focus after passing through a circular aperture. The phase of the wave is the same at all points in the aperture; i.e., the light is coherent. As shown in Figure A-2, the z-axis is chosen to pass through the aperture's center and the geometric focus. The aperture radius $a$, the distance from the aperture's center to the geometric focus $f$, and the angle $\theta'_{zc}$ that is subtended by the aperture's radius at the geometric focus are shown in the figure. The angle $\theta'_{zc}$ is the focusing element's exit angle.

### RAYLEIGH RESOLUTION CRITERION

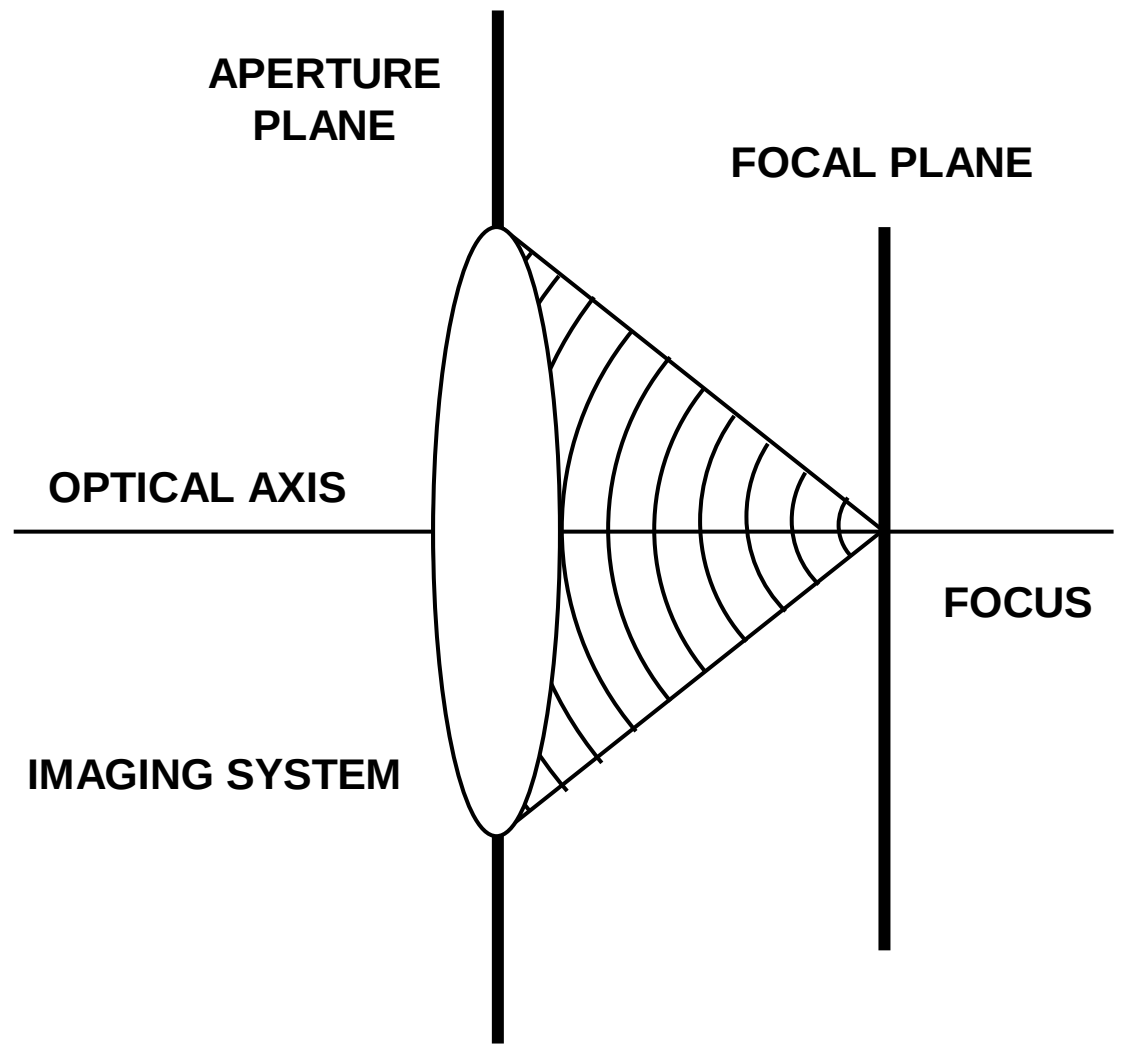

Figure A-1. Apertured converging spherical wave.

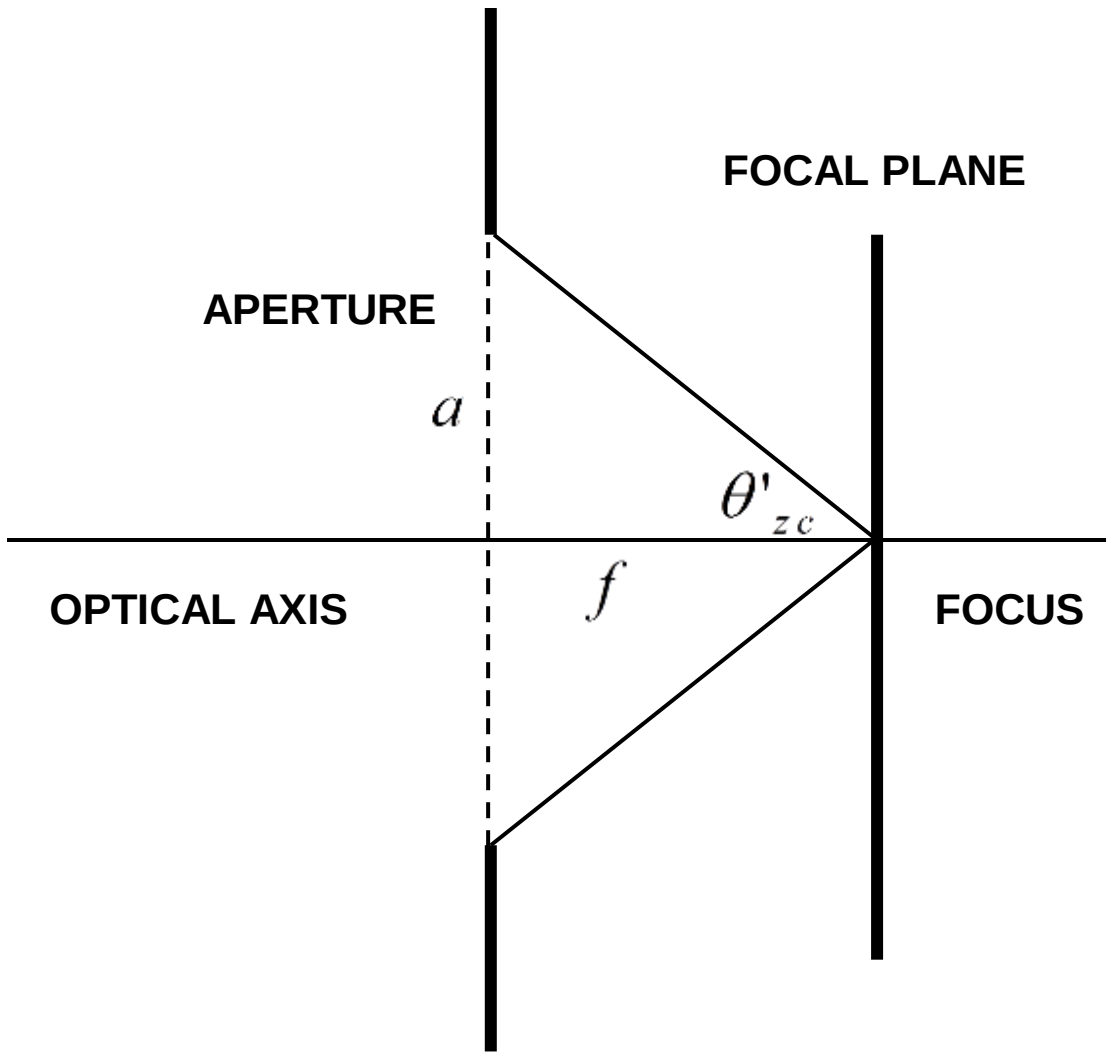

Figure A-2. Geometric parameters.

A three-dimensional distribution of light that is dominated by a bright central region surrounded by a set of alternating dark and bright three-dimensional ovals envelops the focal point[23]. A typical image consists of many of these distributions of light, some of which may (and probably do) overlap. The center-to-center separation (in the focal plane) of two resolved oval-shaped distributions of light defines the Rayleigh resolution criterion. In accord with the Rayleigh resolution criterion, two identical oval-shaped distributions of light are barely resolved (distinguishable) when the center of one of them coincides (on the focal plane) with the first dark oval of the other one.

Let $s_R'$ be the radius (on the focal plane) of the first dark oval associated with a single oval shaped distribution of light. As illustrated in Appendix B, Figure B-2,

$$2\pi \nu'_c s_R' = 1.220\pi \tag{A.1}$$

so that

$$s_R' = \frac{0.61}{\nu'_c} \tag{A.2}$$

when the Rayleigh resolution criterion is satisfied. In accord with equation (8.18)

$$\nu'_c = \frac{(\mathrm{NA})_i}{\lambda_s} \tag{A.3}$$

where $(\mathrm{NA})_i$ is the image side numerical aperture of the focusing element and $\lambda_s$ is the wavelength (in vacuum) of the light used. Substitution of equation (A.3) into equation (A.2) leads to

$$s_R' = \frac{0.61\lambda_s}{(\mathrm{NA})_i} \tag{A.4}$$

readily. The center-to-center separation (on the focal plane) of the two identical oval-shaped distributions of light when they are barely resolved is given by

$$\delta_R' = 2\, s_R' \tag{A.5}$$

Consequently, two points are resolved (barely or otherwise) when the image plane Rayleigh resolution criterion

$$\delta_R' \geq \frac{1.22\lambda_S}{(\mathrm{NA})_i} \tag{A.6}$$

is satisfied.

The projection of the image side Rayleigh resolution criterion through the imaging system onto the object plane is the object side Rayleigh resolution criterion. Consequently, the Rayleigh resolution criteria are given by

$$\begin{pmatrix} \delta_R \\ \delta_R' \end{pmatrix} \geq \begin{pmatrix} \dfrac{1.22\lambda_S}{(\mathrm{NA})_o} \\ \dfrac{1.22\lambda_S}{(\mathrm{NA})_i} \end{pmatrix} \tag{A.7}$$

for both the object plane (unprimed) and the image plane (primed).

Rayleigh resolution criteria are commonly[24] applied far outside their very restricted range[25] of validity. Such application has led to heroic attempts to use ever decreasing wavelengths of light and increasing numerical apertures to improve resolution. This approach has met with a measure of success without a true understanding of the physics involved. The approach constitutes a fight against nature.

# APPENDIX B. CIRCULARLY SYMMETRIC IMAGING SYSTEMS

### INTRODUCTION

Most imaging systems are circularly symmetric. A change of variables from rectangular coordinates to plane polar coordinates in both the *xy*-plane and the $\nu_x\nu_y$-plane is advantageous for treating circularly symmetric imaging systems. Thus

$$\begin{pmatrix} x \\ y \end{pmatrix} = \begin{pmatrix} r\cos\mu \\ r\sin\mu \end{pmatrix} \quad \text{(B.1)}$$

and

$$\begin{pmatrix} \nu_x \\ \nu_y \end{pmatrix} = \begin{pmatrix} \bar{\nu}\cos\phi \\ \bar{\nu}\sin\phi \end{pmatrix} \quad \text{(B.2)}$$

can be appropriately introduced. In addition

$$\sigma(x, y; \lambda) = g(r, \mu; \lambda) \quad \text{(B.3)}$$

and

$$\tilde{\sigma}(\nu_x, \nu_y; \lambda) = \tilde{g}(\bar{\nu}, \phi; \lambda) \quad \text{(B.4)}$$

can be introduced to express the imaging system's coherent impulse response and coherent transfer function, respectively, in polar coordinates.

The Jacobians linked to the transformation from rectangular to polar coordinates are

$$J(r, \mu) = \begin{vmatrix} \cos\mu & -r\sin\mu \\ \sin\mu & r\cos\mu \end{vmatrix} \quad \text{(B.5)}$$

and

$$J(\bar{\nu}, \phi) = \begin{vmatrix} \cos\phi & -\bar{\nu}\sin\phi \\ \sin\phi & \bar{\nu}\cos\varphi \end{vmatrix} \quad \text{(B.6)}$$

which reduce to

$$J(r, \mu) = r \quad \text{(B.7)}$$

and

$$J\left(\bar{\nu},\phi\right)=\bar{\nu} \tag{B.8}$$

respectively. Consequently

$$dy\,dx=r\,dr\,d\mu \tag{B.9}$$

and

$$d\nu_y\;d\nu_x=\bar{\nu}\,d\bar{\nu}\,d\phi \tag{B.10}$$

follow.

Substitution of the foregoing equations into equations (7.1) and (7.2), as appropriate, leads to

$$g\left(r,\mu;\lambda\right)=\int_0^\infty\int_0^{2\pi}\tilde{g}\left(\bar{\nu},\phi;\lambda\right)\exp\left[i2\pi\,r\,\bar{\nu}\left(\cos\mu\cos\phi+\sin\mu\sin\phi\right)\right]\bar{\nu}\,d\bar{\nu}\,d\phi \tag{B.11}$$

and

$$\tilde{g}\left(\bar{\nu},\phi;\lambda\right)=\int_0^\infty\int_0^{2\pi}g\left(r,\mu;\lambda\right)\exp\left[-i2\pi\,r\,\bar{\nu}\left(\cos\mu\cos\phi+\sin\mu\sin\phi\right)\right]r\,dr\,d\mu \tag{B.12}$$

respectively. After introducing the trigonometric identity

$$\cos\mu\cos\phi+\sin\mu\sin\phi=\cos\left(\mu-\phi\right) \tag{B.13}$$

equations (B.11) and (B.12) can be written as

$$g\left(r,\mu;\lambda\right)=\int_0^\infty\int_0^{2\pi}\tilde{g}\left(\bar{\nu},\phi;\lambda\right)\exp\left[i2\pi\,r\,\bar{\nu}\cos\left(\mu-\phi\right)\right]\bar{\nu}\,d\bar{\nu}\,d\phi \tag{B.14}$$

and

$$\tilde{g}\left(\bar{\nu},\phi;\lambda\right)=\int_0^\infty\int_0^{2\pi}g\left(r,\mu;\lambda\right)\exp\left[-i2\pi\,r\,\bar{\nu}\cos\left(\mu-\phi\right)\right]r\,dr\,d\mu \tag{B.15}$$

respectively.

Integral representations of the Bessel function of the first kind of order zero[26], i.e.,

$$J_0\left(-2\pi\,r\,\bar{\nu}\right)=\frac{1}{2\pi}\int_0^{2\pi}\exp\left[i2\pi\,r\,\bar{\nu}\cos\left(\mu-\phi\right)\right]d\phi \tag{B.16}$$

and

$$J_0\left(2\pi r\bar{v}\right)=\frac{1}{2\pi}\int_0^{2\pi}\exp\left[-i2\pi r\bar{v}\cos\left(\mu-\phi\right)\right]d\mu \tag{B.17}$$

can now be advantageously introduced. Equations (B.16) and (B.17) can be rearranged to yield

$$\int_0^{2\pi}\exp\left[i2\pi r\bar{v}\cos\left(\mu-\phi\right)\right]d\phi=2\pi\, J_0\left(-2\pi r\bar{v}\right) \tag{B.18}$$

and

$$\int_0^{2\pi}\exp\left[-i2\pi r\bar{v}\cos\left(\mu-\phi\right)\right]d\mu=2\pi\, J_0\left(2\pi r\bar{v}\right) \tag{B.19}$$

respectively.

Substitution of equation (B.18) into equation (B.14) leads to

$$g\left(r,\mu;\lambda\right)=2\pi\int_0^{\infty}\tilde{g}\left(\bar{v},\phi;\lambda\right)J_0\left(-2\pi r\bar{v}\right)\bar{v}\,d\bar{v} \tag{B.20}$$

while

$$\tilde{g}\left(\bar{v},\phi;\lambda\right)=2\pi\int_0^{\infty}g\left(r,\mu;\lambda\right)J_0\left(2\pi r\bar{v}\right)r\,dr \tag{B.21}$$

can be obtained after substituting equation (B.19) into equation (B.15). The Bessel function of the first kind of order zero is an even function. Consequently

$$J_0\left(-2\pi r\bar{v}\right)=J_0\left(2\pi r\bar{v}\right) \tag{B.22}$$

can be substituted into equation (B.20) to obtain

$$g\left(r,\mu;\lambda\right)=2\pi\int_0^{\infty}\tilde{g}\left(\bar{v},\phi;\lambda\right)J_0\left(2\pi r\bar{v}\right)\bar{v}\,d\bar{v} \tag{B.23}$$

trivially.

### CIRCULARLY SYMMETRIC COHERENT TRANSFER FUNCTION

Coherent transfer functions which are circularly symmetric are necessarily independent of $\phi$; consequently

$$\tilde{g}\left(\bar{\nu},\phi;\lambda\right)=\tilde{h}\left(\bar{\nu};\lambda\right) \tag{B.24}$$

can be appropriately introduced. Substitution of equation (B.24) into equation (B.23) yields

$$g\left(r,\mu;\lambda\right)=2\pi\int_0^\infty \tilde{h}\left(\bar{\nu};\lambda\right)J_0\left(2\pi r\bar{\nu}\right)\bar{\nu}\,d\bar{\nu} \tag{B.25}$$

The right hand side of equation (B.25) is independent of $\mu$. Accordingly

$$g\left(r,\mu;\lambda\right)=h\left(r;\lambda\right) \tag{B.26}$$

can be introduced to reflect the fact that the left hand side of equation (B.25) is necessarily also independent of $\mu$. Substitution of equation (B.26) into equation (B.25) yields

$$h\left(r;\lambda\right)=2\pi\int_0^\infty \tilde{h}\left(\bar{\nu};\lambda\right)J_0\left(2\pi r\bar{\nu}\right)\bar{\nu}\,d\bar{\nu} \tag{B.27}$$

Substitution of equations (B.24) and (B.26) into equation (B.21) yields

$$\tilde{h}\left(\bar{\nu};\lambda\right)=2\pi\int_0^\infty h\left(r;\lambda\right)J_0\left(2\pi r\bar{\nu}\right)\bar{\nu}\,d\bar{\nu} \tag{B.28}$$

The circularly symmetric coherent transfer function $\tilde{h}\left(\bar{\nu};\lambda\right)$ and the circularly symmetric coherent impulse response $h\left(r;\lambda\right)$ are related to one another by means of equations (B.27) and (B.28).

Substitution of equation (B.24) into equation (B.4) and then substituting the result into equation (7.3) leads to

$$\tilde{h}\left(\bar{\nu};\lambda\right)=\begin{cases}1 & \text{inside the passband}\\ 0 & \text{outside the passband}\end{cases} \tag{B.29}$$

which is the circularly symmetric coherent transfer function. Equation (B.29) can be written as

$$\tilde{h}\left(\bar{\nu};\lambda\right)=\begin{cases}1 & \bar{\nu}\le\nu_c\\ 0 & \bar{\nu}>\nu_c\end{cases} \tag{B.30}$$

where $\nu_c$ is the spatial frequency cutoff of the imaging system.

## CIRCULARLY SYMMETRIC COHERENT IMPULSE RESPONSE

Equation (B.27) relates the circularly symmetric coherent impulse response $h(r;\lambda)$ to the circularly symmetric coherent transfer function $\tilde{h}(\bar{v};\lambda)$. Substitution of the circularly symmetric transfer function given by equation (B.30) into equation (B.27) Leads to

$$h(r;\lambda) = 2\pi \int_0^{v_c} \bar{v}\, J_0(2\pi r \bar{v};\lambda)\, d\bar{v} \qquad \text{(B.31)}$$

readily. Equation (B.31) can be rewritten as

$$h(r;\lambda) = \frac{1}{2\pi r^2} \int_0^{v_c} 2\pi r \bar{v}\, J_0(2\pi r \bar{v};\lambda)\, 2\pi r\, d\bar{v} \qquad \text{(B.32)}$$

advantageously. Many authors, together with Hecht[27] and Born and Wolf[28], treat circularly symmetric coherent impulse response functions and related topics.

Let $J_1(2\pi r \bar{v};\lambda)$ be the Bessel function of the first kind of order one. The recursion relation

$$\left(\frac{1}{2\pi r}\right) \frac{d}{d\bar{v}} \left[2\pi r \bar{v}\, J_1(2\pi r \bar{v};\lambda)\right] = 2\pi r \bar{v}\, J_o(2\pi r \bar{v};\lambda) \qquad \text{(B.33)}$$

is introduced and discussed by many authors, including Boas[29], Hecht[30], and Born and Wolf[31]. Equation (B.33) can be rearranged and written as

$$d\left[2\pi r \bar{v}\, J_1(2\pi r \bar{v};\lambda)\right] = 2\pi r \bar{v}\, J_o(2\pi r \bar{v};\lambda)\, 2\pi r\, d\bar{v} \qquad \text{(B.34)}$$

readily. Substitution of equation (B.34) into equation (B.32) yields

$$h(r;\lambda) = \frac{1}{2\pi r^2} \int_0^{v_c} d\left[2\pi r \bar{v}\, J_1(2\pi r \bar{v};\lambda)\right] \qquad \text{(B.35)}$$

immediately. Upon evaluation, equation (B.35) yields

$$h(r;\lambda) = \frac{1}{2\pi r^2} \left[2\pi v_c r\, J_1(2\pi v_c r;\lambda)\right] \qquad \text{(B.36)}$$

in a straightforward manner. Equation (B.36) can be simplified to obtain

$$h(r;\lambda) = \pi {v_c}^2 \left[\frac{2 J_1(2\pi v_c r;\lambda)}{2\pi v_c r}\right] \qquad \text{(B.37)}$$

readily. Equation (B.37) describes the coherent impulse response for a circularly symmetric imaging system.

The relative probability density associated with the impulse response provided by equation (B.37) is given by

$$I(r;\lambda) = h^*(r;\lambda)\, h(r;\lambda) \tag{B.38}$$

Substitution of equation (B.37) into equation (B.38) yields

$$I(r;\lambda) = \left(\pi \nu_c^{\,2}\right)^2 \left[ \frac{2J_1(2\pi \nu_c r;\lambda)}{2\pi \nu_c r} \right]^2 \tag{B.39}$$

trivially. This result is known as the Airy formula.

## AIRY DISK

A graph of the standardized Airy formula, i.e.

$$\frac{I(r;\lambda)}{\left(\pi \nu_c^{\,2}\right)^2} = \left[ \frac{2J_1(2\pi \nu_c r;\lambda)}{2\pi \nu_c r} \right]^2 \tag{B.40}$$

is provided in Figure B-1. As illustrated in Figure B-1, the optical real image linked to the standardized Airy formula is an Airy pattern.

As shown in Figures B-1 and B-2, an Airy pattern consists of a central bright region that is surrounded by a number of much fainter rings. The central bright region is known as the Airy disk associated with the imaging system.

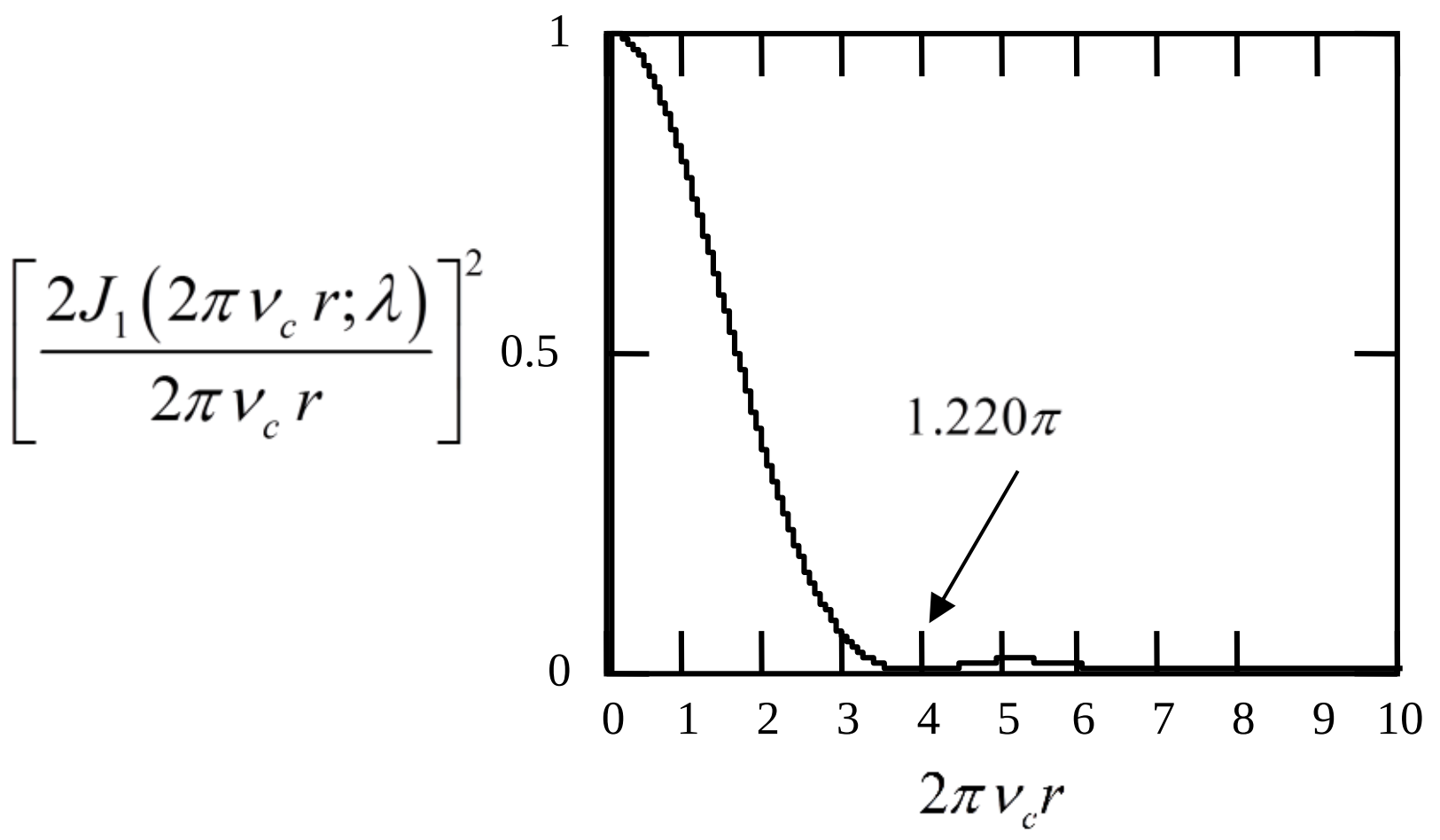

Figure B-1. Standardized Airy formula.

An Airy disk is bounded by a dark ring that exists at a location that corresponds to the first zero of the Bessel function $J_1\left(2\pi r \nu_c ; \lambda\right)$[32]. As indicated in Figure B-1, this occurs when the value of the independent variable in equation (B.40) has the value $1.220\pi$, i.e. when

$$2\pi \nu_c r_0 = 1.220\pi \quad \text{(B.41)}$$

where the radius of the Airy disk

$$r_0 = \frac{0.610}{\nu_c} \quad \text{(B.42)}$$

has been introduced. Correspondingly,

$$\delta = \frac{1.220}{\nu_c} \quad \text{(B.43)}$$

is the diameter of the Airy disk.

**AIRY DISK DIAMETER**

Equation (B.43) provides the Airy disk diameter in the region where unprimed parameters are used (the region between the object plane and the imaging system). When the region where primed parameters are used (the region between the imaging system and the image plane) is also included, the Airy disk diameter is given by

$$\begin{pmatrix} \delta \\ \delta' \end{pmatrix} = \begin{pmatrix} \dfrac{1.220}{\nu_c} \\ \dfrac{1.220}{\nu'_c} \end{pmatrix} \tag{B.44}$$

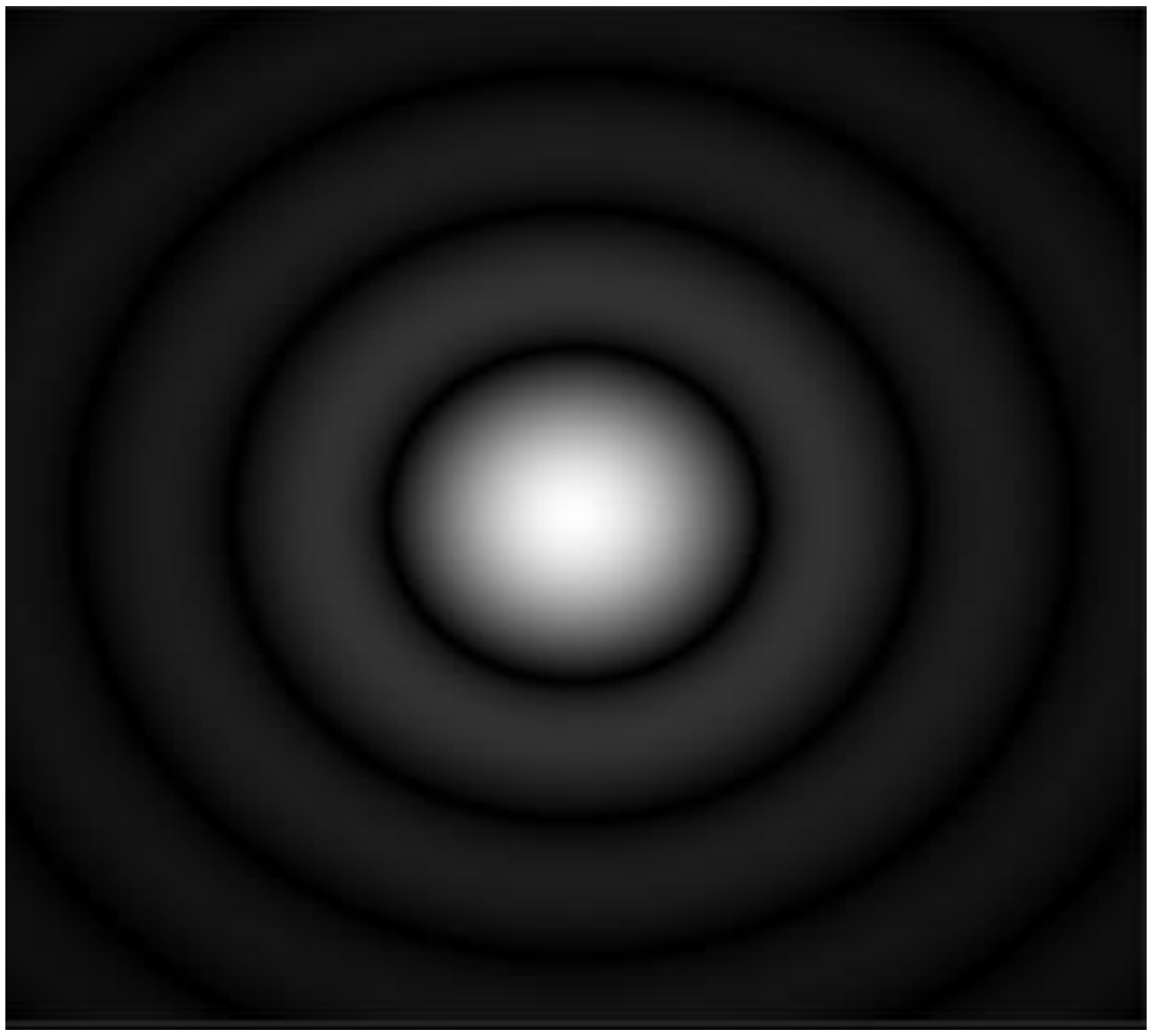

Figure B-2. Airy pattern. Source: Wikipedia, *Airy Disk*.

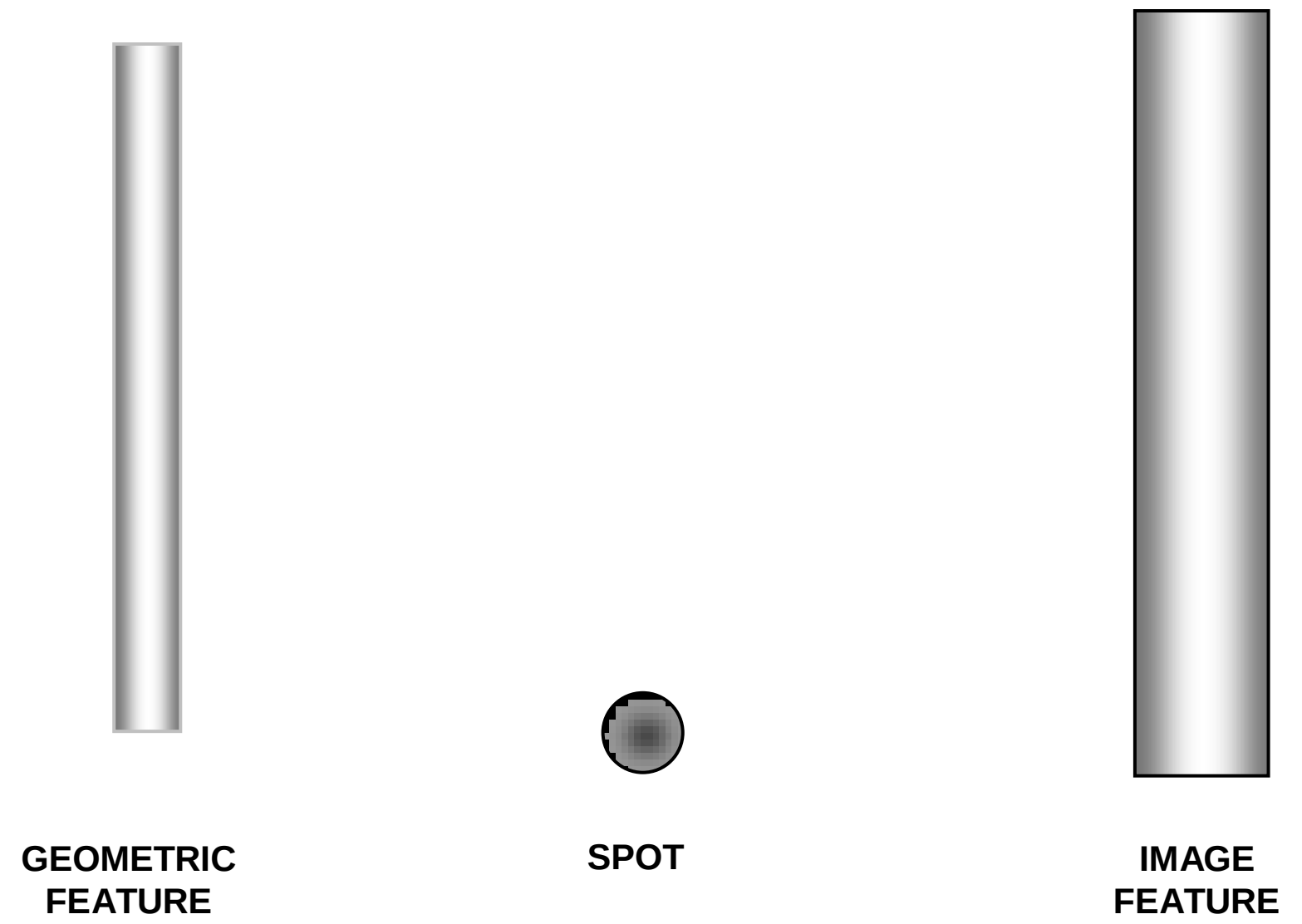

Figure B-3. Conventional image feature dimensions are increased by the spot diameter.

After substituting equation (8.18) into equation (B.44)

$$\begin{pmatrix} \delta \\ \delta' \end{pmatrix} = \begin{pmatrix} \dfrac{1.220\lambda_S}{(\mathrm{NA})_o} \\ \dfrac{1.220\lambda_S}{(\mathrm{NA})_i} \end{pmatrix} \tag{B.45}$$

results. The diameter of the Airy disk, as expressed in terms of primed parameters, is often referred to as the imaging system's spot diameter.

Dimensions of image features that are formed by conventional means are larger than their geometrically defined size. As indicated in Figure B-3, the linear dimensions of such an image feature are increased in size by an amount equivalent to the imaging system's spot diameter.

## REFERENCES

[1] Greyson Gilson, "Unified Theory of Wave-Particle Duality, Derivation of the Schrödinger Equations, and Quantum Diffraction," arXiv:1103.1922 (18 June 2014).
[2] Eugene Hecht, *Optics*, (4th Ed.) p. 52 (Addison Wesley, San Francisco, 2002).
[3] Greyson Gilson, "Unified Theory of Wave-Particle Duality, Derivation of the Schrödinger Equations, and Quantum Diffraction," arXiv:1103.1922 (18 June 2014).
[4] David Halliday and Robert Resnick, *Physics*, p. 986 (John Wiley & Sons, Inc., New York, 1960).
[5] John David Jackson, *Classical Electrodynamics*, pp. 202-203 (John Wiley & Sons, Inc., New York, 1967).
[6] David Halliday and Robert Resnick, *Physics*, p. 986 (John Wiley & Sons, Inc., New York, 1960).
[7] Eugene Hecht, *Optics* (4th Ed.) p. 8, pp. 51-56, and pp.137-139 (Addison Wesley, San Francisco, 2002).
[8] A. P. French and Edwin F. Taylor, *An Introduction to Quantum Physics*, pp. 87-95 (W. W. Norton & Company Inc., New York 1978).
[9] Robert J. Collier, Christoph B. Burkhardt and Lawrence H. Lin, *Optical Holography*, p. 81-83 (Academic Press, Inc., Orlando, 1971).
[10] Jack D. Gaskill, *Linear Systems, Fourier Transforms, and Optics*, pp. 454-471 (John Wiley & Sons, Inc., New York, 1978).
[11] W. Thomas Cathey, *Optical Information Processing and Holography*, pp. 114-116 (John Wiley & Sons, Inc., New York, 1974).
[12] Joseph W. Goodman, *Introduction to Fourier Optics*, pp. 102-105 (McGraw-Hill, Inc., New York, 1968).
[13] Eugene Hecht, *Optics*, (4th Ed.) p. 445, p. 456, and p. 489 (Addison Wesley, San Francisco, 2002).
[14] Greyson Gilson, "Unified Theory of Wave-Particle Duality, Derivation of the Schrödinger Equations, and Quantum Diffraction," arXiv:1103.1922 (18 June 2014).
[15] Greyson Gilson, "Unified Theory of Wave-Particle Duality, Derivation of the Schrödinger Equations, and Quantum Diffraction," arXiv:1103.1922 (18 June 2014).
[16] Greyson Gilson, "Unified Theory of Wave-Particle Duality, Derivation of the Schrödinger Equations, and Quantum Diffraction," arXiv: 1103.1922 (18 June 2014).
[17] Greyson Gilson, "Unified Theory of Wave-Particle Duality, Derivation of the Schrödinger Equations, and Quantum Diffraction," arXiv:1103.1922 (18 June 2014).
[18] Greyson Gilson, "Unified Theory of Wave-Particle Duality, Derivation of the Schrödinger Equations, and Quantum Diffraction," arXiv:1103.1922 (18 June 2014).
[19] Jack D. Gaskill, *Linear Systems, Fourier Transforms, and Optics*, pp. 456-461 (John Wiley & Sons, Inc., New York, 1978).
[20] Joseph W. Goodman, *Introduction to Fourier Optics*, pp. 110-113 (McGraw-Hill, Inc., New York, 1968).
[21] A. J. den Decker and A. van den Bos, *Resolution: a Survey*. J. Opt. Soc. Am. A, Vol 14, No. 3, pp. 547-557. (March 1997).
[22] Chris Mack, *Fundamental Principles of Optical Lithography: The Science of Microfabrication*, p. 105 (John Wiley & Sons, Ltd., Chichester, England, 2007).
[23] Max Born and Emil Wolf, *Principles of Optics* (Sixth Corrected Edition) pp.439-441 (Pergamon Press, Oxford, 1980).
[24] Chris Mack, *Fundamental Principles of Optical Lithography: The Science of Microfabrication* (John Wiley & Sons, Ltd., Chichester, England, 2007).
[25] Colin J.R. Sheppard, *The Optics of Microscopy*, J. Opt. A: Pure Appl. Opt. **9,** pp. S1–S6 (2007).
[26] Frank Bowman, *Introduction to Bessel Functions*, p. 57 (Dover Publications Inc., New York, 1958).
[27] Eugene Hecht, *Optics*, (4th Ed.) pp. 467-471 (Addison Wesley, San Francisco, 2002).
[28] Max Born and Emil Wolf, *Principles of Optics* (Sixth Corrected Edition) pp. 395-398. (Pergamon Press, Oxford, 1980).
[29] Mary L. Boas, *Mathematical Methods in the Physical Sciences*, p. 564. (John Wiley & Sons, Inc., New York, 1966).
[30] Eugene Hecht, *Optics*, (4th Ed) p. 468 (Addison Wesley, San Francisco, 2002).
[31] Max Born and Emil Wolf, *Principles of Optics* (Sixth Corrected Edition), p. 396 (Pergamon Press, Oxford, 1980).
[32] Max Born and Emil Wolf, *Principles of Optics* (Sixth Corrected Edition), p. 397 (Pergamon Press, Oxford, 1980).